\documentclass[preprint2]{aastex}
\usepackage{natbib}
\usepackage[caption=false]{subfig}
\usepackage{dblfloatfix}
\usepackage{float}
\shorttitle{Impulsive Flare Footpoint EMDs}
\shortauthors{Graham et al.}

\begin{document}


\title{The emission measure distribution of impulsive phase flare footpoints}


\author{D.R. Graham\altaffilmark{1}, I.G. Hannah\altaffilmark{1}, L.
Fletcher\altaffilmark{1} and R. O. Milligan\altaffilmark{2}}
\affil{School of Physics and Astronomy, University of Glasgow, G12 8QQ, UK.}
\affil{Astrophysics Research Centre, School of Mathematics and Physics, Queens University Belfast, Belfast, BT7 1NN, UK.}

\email{d.graham@astro.gla.ac.uk}


%%\altaffiltext{1}{School of Physics and Astronomy, University of
%%Glasgow,Glasgow, G12 8QQ, UK.}
%%\altaffiltext{2}{Queens University Belfast, Belfast, BT7 1NN, UK.}


\begin{abstract}
The temperature distribution of the emitting plasma is a crucial constraint when
studying the heating of solar flare footpoints. However, determining this for
impulsive phase footpoints has been difficult in the past due to insufficient
spatial resolution to resolve the footpoints from the loop structures, and a
lack of spectral and temporal coverage.  We use the capabilities of
\emph{Hinode}/EIS to obtain the first emission measure distributions (EMDs) from
impulsive phase footpoints in six flares. Observations with good spectral
coverage were analysed using a regularized inversion method to recover the EMDs.
We find that the EMDs all share a peak temperature of around 8 MK, with lines
formed around this temperature having emission measures peaking between
$10^{28}$ and $10^{29} \rm~cm^{-5}$, indicating a substantial presence of plasma
at very high temperatures within the footpoints. An EMD gradient of EM(T) $\sim
T$ is found in all events. Previous theoretical work on emission measure
gradients shows this 
to be consistent with a scenario in which the deposited flare energy directly
heats only the top layer of the flare chromosphere, while deeper layers are
heated by conduction.

\end{abstract}


\section{Introduction}
A clear observational description of the plasma properties of the lower
atmosphere footpoints of solar flares provides a critical constraint on the
distribution of the flare excess energy in this region, and hence the profile of
flare energy deposition and its possible modes of transport. The emission
measure distribution (EMD) is a way to describe the amount of emitting plasma as
a function of its temperature, and in this paper we present the first
determination of impulsive phase flare footpoint EMDs made using the Extreme
Ultraviolet Imaging Spectrometer \citep[EIS, ][]{2007SoPh..243...19C} onboard
the {\emph{Hinode}} satellite. The temperature coverage of EIS makes it
extremely well suited to studying the properties of footpoints during flares,
where impulsive stage temperatures can be very high - on the order of
8-10~MK, but present in the lower atmosphere during the extreme conditions of a
flare.

There have been several previous studies of the distribution of emission
measures of solar flares. For example, EUV data from the Skylab NRL slitless
spectroheliograph enabled spatial as well as spectral information to be deduced,
though the ``overlappograms'' produced convolved spatial and spectral
information leading to source confusion.  Several flare EMDs from the rise and
decay phases have been published from this instrument. The EMD of
\cite{1977ApJ...217..976D}, from just before the peak of the 0.5-3~\AA~rise
phase of flare SOL1973-09-05T18:32\footnote{we use the solar observation target
identification convention described by \citet{2010SoPh..263....1L}}, which would
correspond roughly to the end of the impulsive phase, had a steep slope up to a
peak at $\log T  \sim 6.9$. There was evidence for emission at Fe {\sc xiv} to 
Fe {\sc xvi} at concentrated footpoints or near the ends of loops, as well as in
loops themselves.

The distribution of emission measures\footnote{A distinction should be made here
between the EMD and {\emph {differential}} emission measure (DEM). The DEM
$\xi(T)$ is the quantity of units $\rm{cm^{-5}K^{-1}}$ derived from an inversion
of the data, e.g. via the method described in Section 4. Integrating the DEM
over a fixed logarithmic temperature interval gives the emission measure as a
function of temperature $EM(T)$ in the more practical units of $\rm{cm^{-5}}$}
found in the impulsive phase of a flare by \citet{1982ApJ...258..835W} in
SOL1973-12-22T00:24 had slope $\int n_e(T)^2 dS \sim T^{0.8}$ from $\log T =
5.4$ up to $\log T = 6.2$, with the suggestion of a decrease above $\log T \sim
6.9$.  Here, $n_e$ is the electron number density and $S$ the distance along the
line of sight. \cite{1984ApJ...281..426W} presented emission measure
distributions from two compact flare sources just after the impulsive peak of
SOL1974-01-21T23:24, one of which had a shallow slope (scaling as $T^{0.6}$) up
to a maximum at $\log T\sim 6 $, and the other had a slope of $3$ up to  $\log T\sim 6.9$. 
\cite{1982ApJ...258..835W} emphasizes that these slopes are different from the
$3/2$ slopes which are often assumed to be characteristic of flares. Other
observations have been obtained during decay phases of flares
\citep{1979ApJ...229..772D,1980ApJ...242.1243S}, and have also tended to attract
more theoretical attention \citep{1979ApJ...232..903M}.

High densities in impulsive-phase sources were also found using the Skylab data,
including in high temperature lines
\citep{1977ApJ...215..329D,1977ApJ...215..652F}. Related observations from the
{\emph{Yohkoh}} satellite \citep{1992Sci...258..618A} have also shown high
temperatures and densities during the impulsive phase, and more specifically
soft X-ray emission from the footpoints \citep{1993ApJ...416L..91M,
1994ApJ...422L..25H, 2004A&A...415..377M}. High footpoint temperature and
density observations have recently been amply confirmed by EIS
\citep[e.g.][]{2011ApJ...740...70M,2011A&A...526A...1D,2011A&A...532A..27G,
2010ApJ...719..213W}.

In this paper, we are able to isolate the footpoints in six flares, and
determine their impulsive phase EMDs using EIS. The $\sim$ 4\arcsec\ spatial
resolution of EIS is sufficient to make a clear identification of footpoints and
loops, based on their morphology at different wavelengths. We identify six
flares having consistent spectral coverage and raster observations during their
impulsive phases. The EMDs recovered are remarkably consistent with one another,
peaking at a temperature of $\log T \sim 6.9$, at a maximum value of
$10^{28}-10^{29}\rm{cm}^{-5}$ and with EMDs below the peak temperature
characterized by $EM(T) \sim T$.

Section \ref{sec:eis} details the EIS data reduction, while an overview of the
flare observations and their selection is found in Section \ref{sec:flareobs}.
Section 4 describes the inputs required in calculating the DEM and the
regularized inversion method used to determine the DEM. The footpoint EMDs are
shown in Section 5 and discussed in depth in Sections 6 and 7.

\section{EIS Observations}\label{sec:eis}
To obtain EMDs of flare footpoint plasma we require observations of flares
during the impulsive (rise) phase at multiple temperatures, ideally in emission
lines given their narrow sensitivity to temperature (see Section
\ref{sec:demtech}). Its spatial, temporal, and spectral capabilities make {\it
Hinode}/EIS well suited for this task. The spectrometer slit can be scanned
across the area of an active region multiple times during the flare, building up
raster ``images'' in many wavelengths simultaneously. Good datasets for flare
footpoint studies are hard to find. Positioning the slit over the small
footpoints (typically 2-5\arcsec) early in the flare is not always possible, and
since March 2008 telemetry from the Hinode spacecraft has been limited,
restricting the temperature sampling.

\begin{figure}[h!]
 \centering
 \includegraphics[width=8cm]{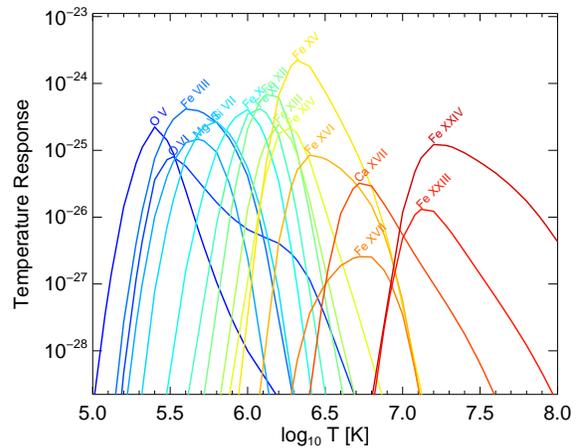}
 \caption{Contribution functions $G(T)$ for emission lines calculated with
coronal abundances, CHIANTI's own ionization equilibrium file, and a
constant density of $10^{11} {\rm cm^{-3}}$}
 \label{fig:goft}
\end{figure}

Flare observations using the EIS raster study CAM\_ARTB\_RHESSI\_b\_2 fulfilled
our requirements and six were selected. These use the 2\arcsec\ slit to scan a
40\arcsec\ x\ 140\arcsec\ area in 3 min 52 s. Around 30 lines are present in
these rasters but we have narrowed the selection down to 15 to best cover the
temperature range (Figure \ref{fig:goft}), and avoid density-sensitive or
optically thick lines. Line details are listed in Table \ref{tab:lines} and
contribution functions in Figure \ref{fig:goft}. All of the data has been
calibrated and fitted using the standard {\sc eis\_prep} and {\sc
eis\_auto\_fit} SolarSoft routines. An exception is the analysis of Ca~{\sc
xvii} which uses the method described below. We also correct for a measured
17\arcsec\ North-South and 1\arcsec\ East-West offset between the two separate
wavelength bands on the instrument's CCD.

\begin{table}[h!]
\caption{Emission lines selected for EMD analysis with rest wavelengths and peak
formation temperatures.}\label{tab:lines}
 \begin{center}
 \begin{tabular}{ccc}
\tableline\tableline

Ion & $\lambda$ (\AA) & $log_{10} T$ (K)\\
\tableline
O {\sc v} & 248.460 & 5.4\\
O {\sc vi} & 184.118 & 5.5\\
Fe {\sc viii} & 185.213 & 5.7\\
Mg {\sc vi} & 268.991 & 5.7\\
Si {\sc vii} & 275.361 & 5.8\\
Fe {\sc x} & 184.537 & 6.1\\
Fe {\sc xi} & 188.216 & 6.2\\
Fe {\sc xii} & 195.119 & 6.2\\
Fe {\sc xiii} & 202.044 & 6.3\\
Fe {\sc xiv} & 274.204 & 6.3\\
Fe {\sc xv} & 284.163 & 6.4\\
Fe {\sc xvi} & 262.976 & 6.4\\
Ca {\sc xvii} & 192.853 & 6.8\\
Fe {\sc xxiii} & 263.766 & 7.2\\
Fe {\sc xxiv} & 192.028 & 7.2\\
\tableline
\end{tabular}
\end{center}
\end{table}

We extract for each emission line the fitted integrated line intensity averaged
over a 2\arcsec\ $\times$\ 3\arcsec\ area around a footpoint (1$\times$\ 3
pixels) centred on the pixel brightest in Fe~{\sc viii}. The strongest footpoint
emission in these events mostly appears over one slit position 2\arcsec\ wide.
To bin further in the $x$-direction would sample too much of the surrounding
area, thus our binning accounts for the spatial extent of the footpoint and
covers some offset between wavelengths.

A pre-flare background was not subtracted from the footpoint emission. This is
allows us to make consistent comparisons with the EMDs described in the
introduction, and the theoretical work discussed in Section
\ref{sec:discussion}; which treats emission from the entire emitting column, not
only the flare excess.

A number of the lines selected are blended by neighboring transitions, however
the true intensities can be recovered using fitting techniques and other
observed lines within the raster. The CHIANTI v7.0 atomic database
\citep{1997A&AS..125..149D, 2012ApJ...744...99L} is used to identify lines
contributing to the measured line profiles.

The low temperature Fe {\sc viii}~185.213 line has two reported blends of
Ni {\sc xxiv}~185.166 and Ni {\sc xvi} 185.230. Fitting with three Gaussians
reveals small 10-20\% blue wing contributions of Ni {\sc xxiv}~185.166 around, but
not necessarily within, the footpoints. In the red wing CHIANTI predicts Ni {\sc
xvi} 185.230 to be the stronger contribution, however, no obvious third Gaussian
could be seen from our fitting and so Ni {\sc xvi} should not have a significant effect on the Fe
{\sc viii} intensity.

%Fe {\sc xi}~188.216 is self-blended with another Fe~{\sc xi} line at 188.299
%which is easily resolved and removed with a double Gaussian fit.

Fe {\sc xxiv} 192.028 normally dominates in flare loop conditions and
contributions from blends of Fe {\sc viii} and Fe {\sc xi} are small. This work
concentrates on an earlier phase in the flare where the Fe {\sc xxiv} emission
will be much fainter, therefore a significant Fe {\sc xi} 192.021
contribution must be removed. We do this via the method described in
\cite{2011A&A...526A...1D}. The observed Fe~{\sc xi} 201.734 intensity forms a
known ratio with $\lambda$192.021, this can then be used to estimate the
$\lambda192.021$ intensity in the footpoint. We find the ratio of 192.021/201.734 in background moss/active regions to be 0.43. The Fe~{\sc viii} 192.004 intensity is predicted by CHIANTI to be around 10\% of the Fe {\sc xi} line and is almost negligible. Accounting for these contributions removes most of the active-region emission (seen in Figure \ref{fig:hotcold}) leaving a mean background level lower than the line uncertainties (see Section \ref{sec:errors} on DEM uncertainties).

Ca {\sc xvii} 192.853 forms part of a complex grouping of six O {\sc v} and two
Fe {\sc xi} lines, therefore interpretation of the line is challenging but is
well documented \citep{2009ApJ...697.1956K, 2010A&A...521A..21O,
2011A&A...526A...1D} and we have adopted the approach in
\citet{2009ApJ...697.1956K} to estimate the Ca~{\sc xvii} intensity. First the
two Fe {\sc xi} 192.813/192.901 intensities are estimated using known ratios to
Fe~{\sc xi} 188.216. In most of the footpoints the strong density sensitive O
{\sc v} 192.904 line can be resolved and the other five O {\sc v} lines are
estimated from CHIANTI v7.0 using a fixed density of $10^{11} {\rm cm^{-3}}$.

\section{Flare Observations and Footpoint Selection}\label{sec:flareobs}
\begin{figure*}[t]
 \centering
 \subfloat{{(a)}\includegraphics[width=5cm]{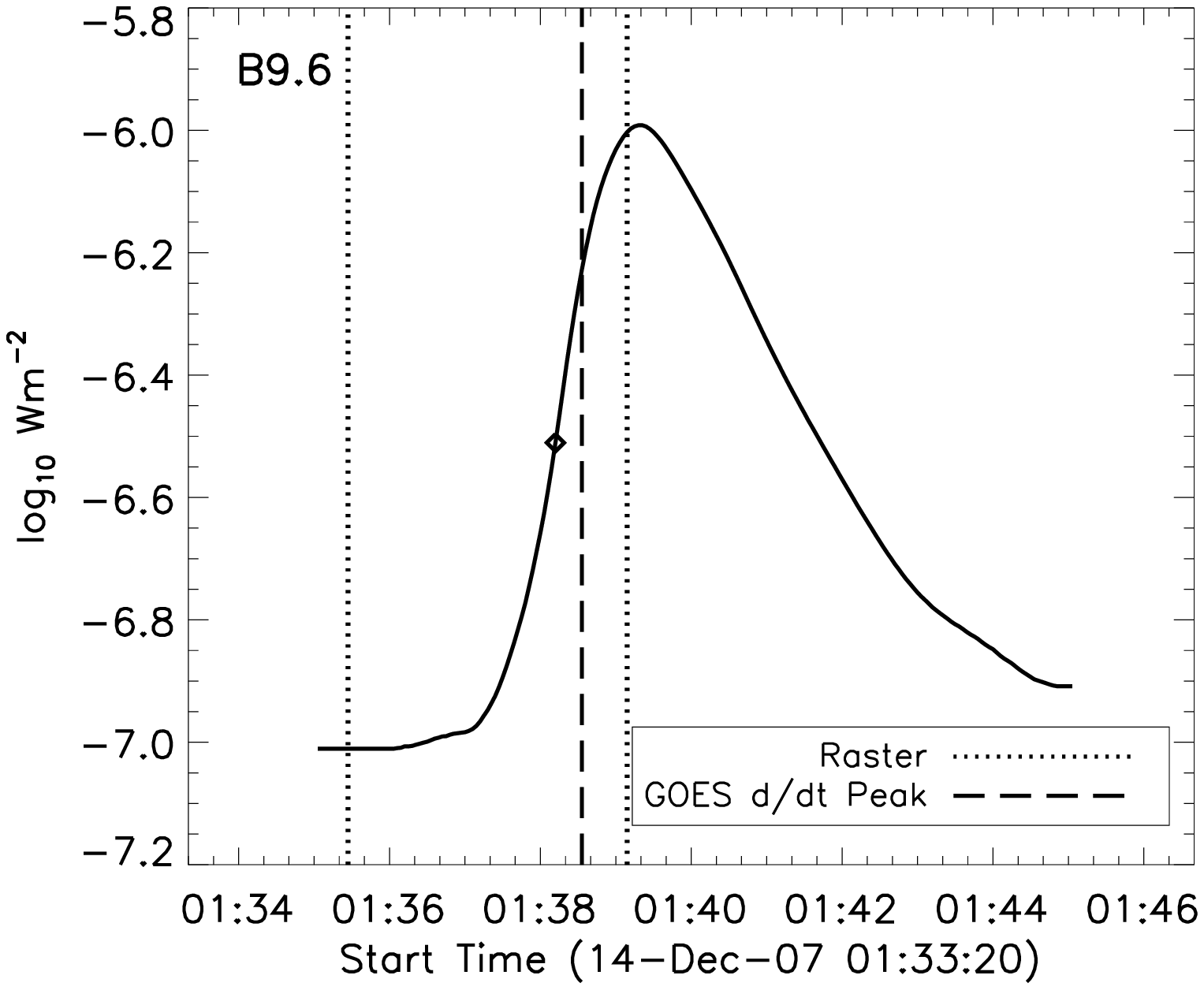}}
 \subfloat{{(b)}\includegraphics[width=5cm]{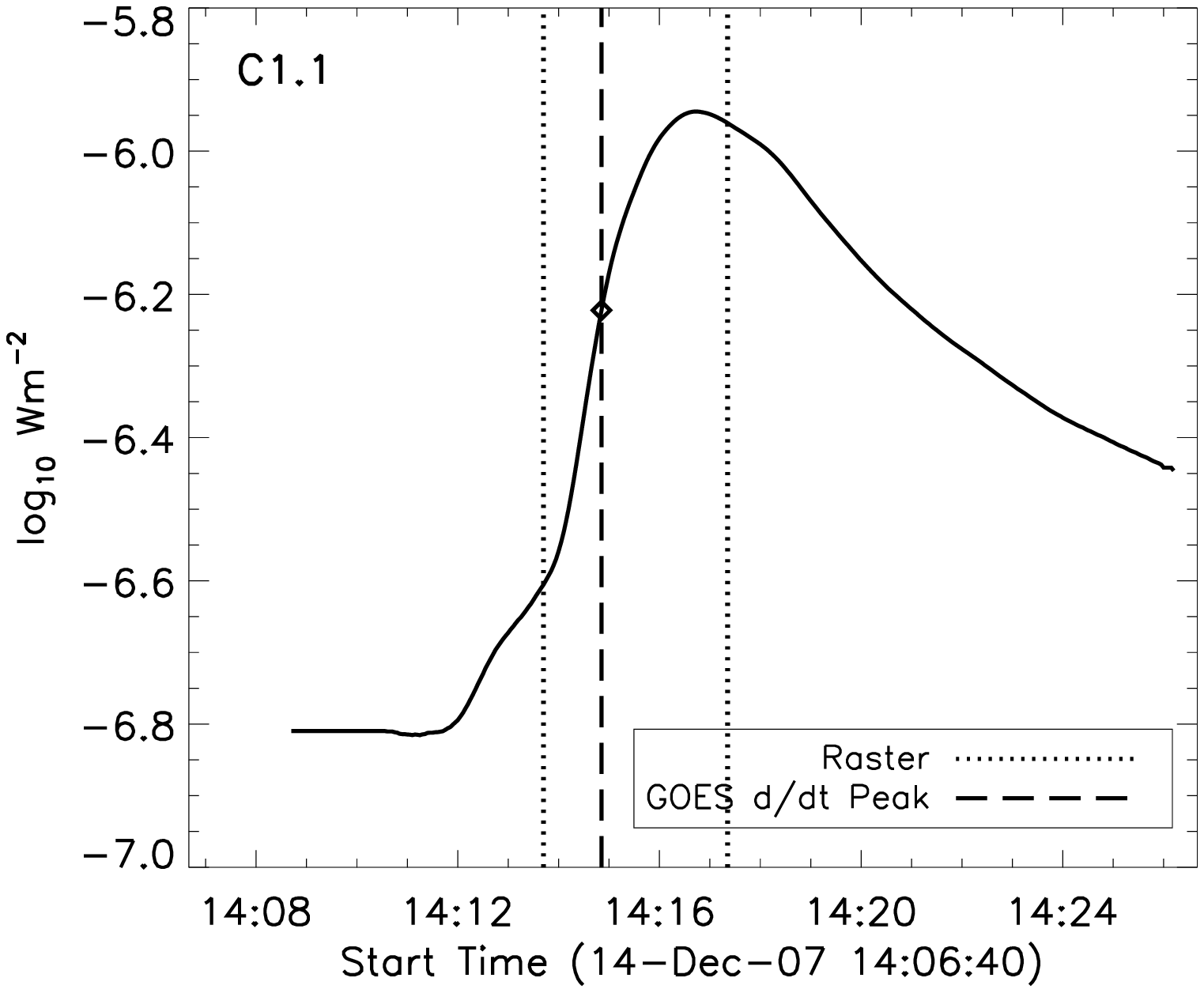}}
 \subfloat{{(c)}\includegraphics[width=5cm]{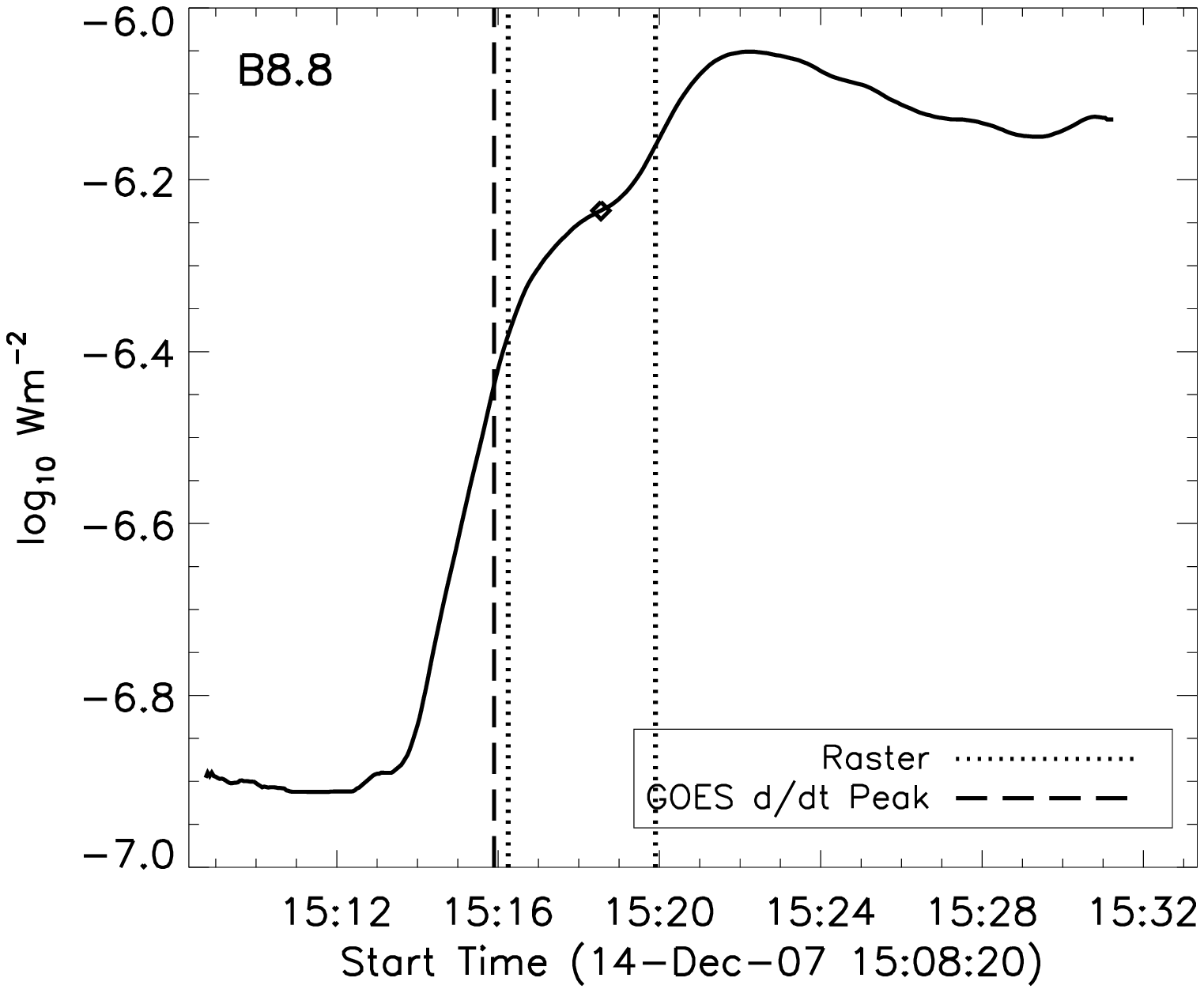}}

 \subfloat{{(d)}\includegraphics[width=5cm]{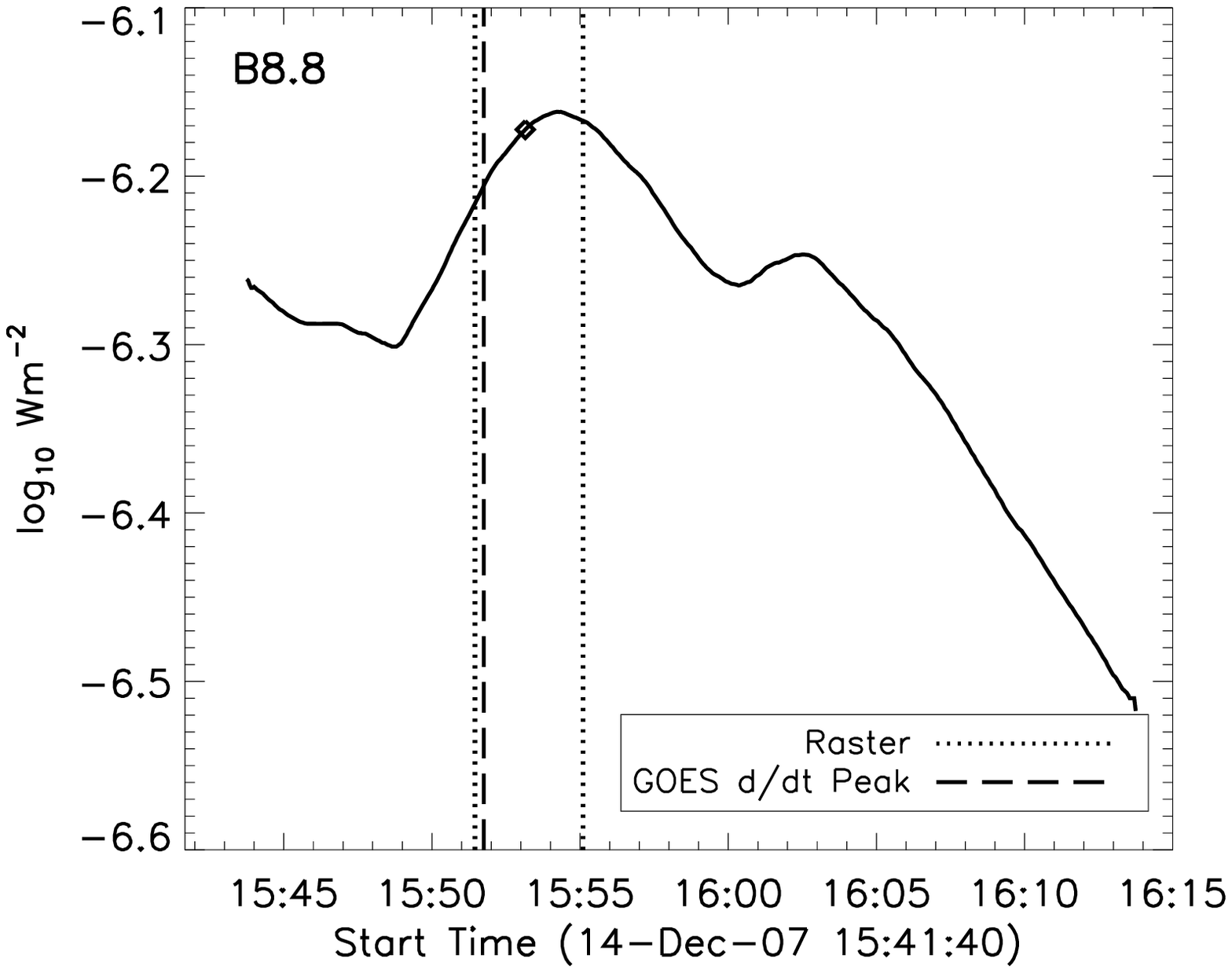}}
 \subfloat{{(e)}\includegraphics[width=5cm]{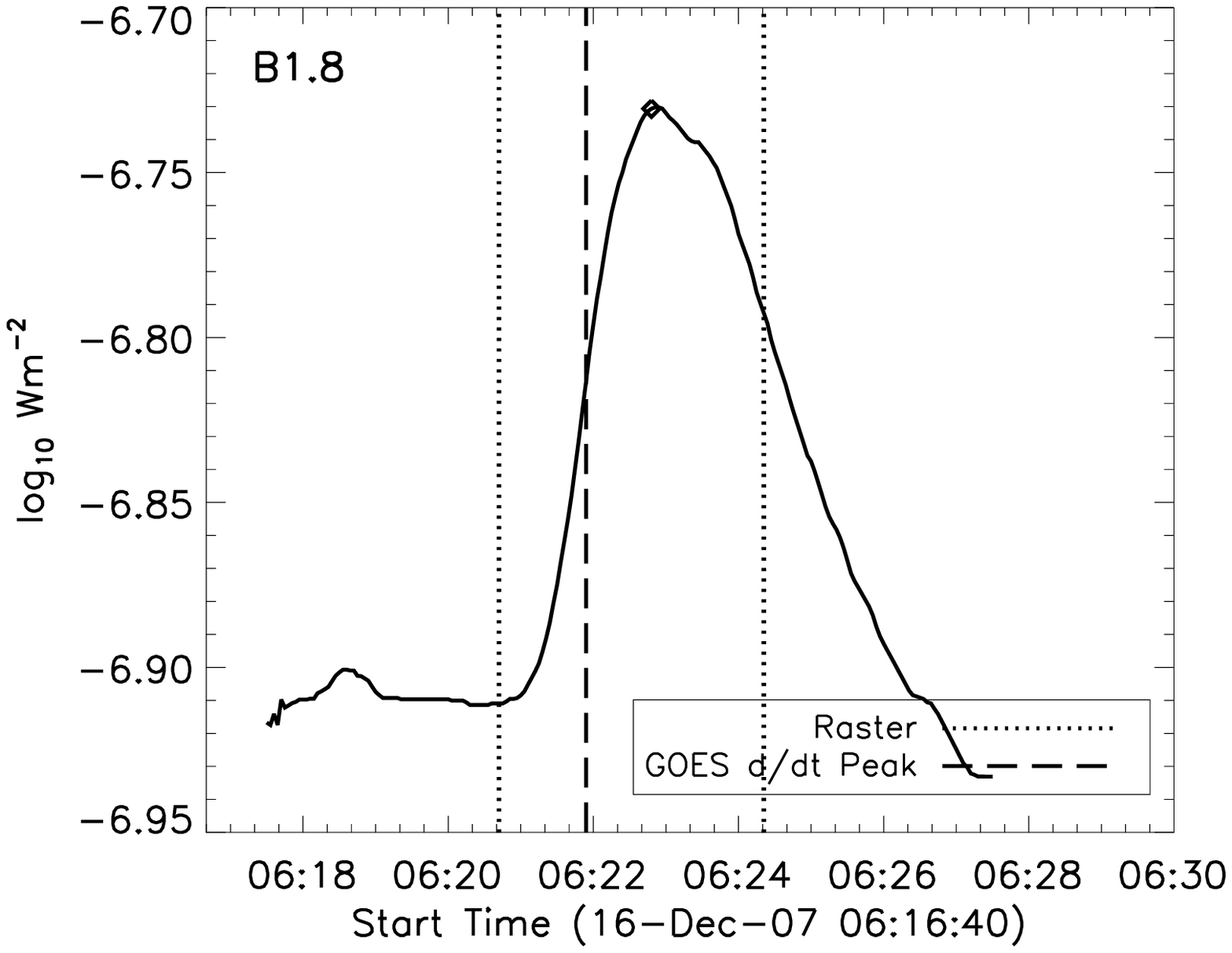}}
 \subfloat{{(f)}\includegraphics[width=5cm]{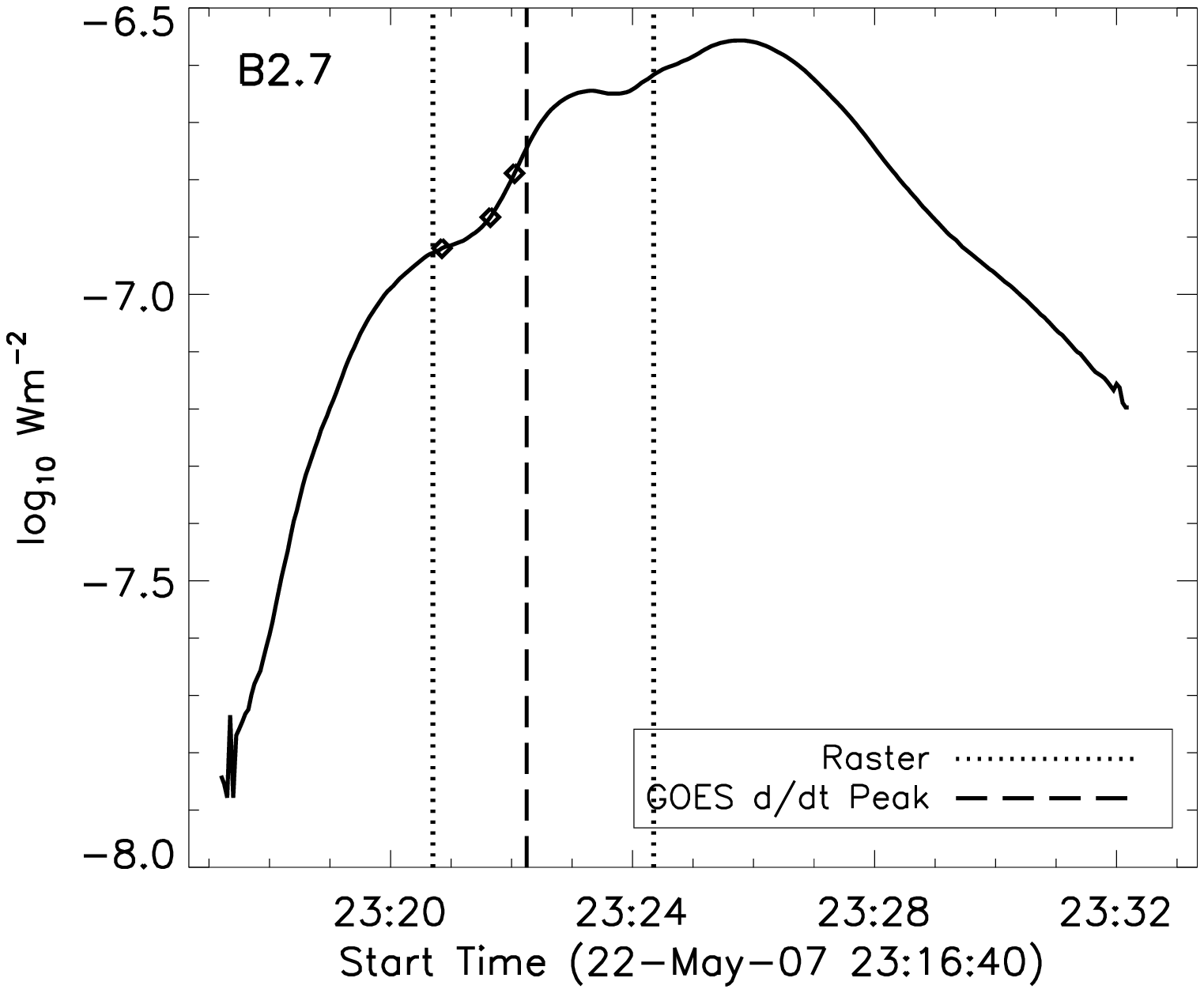}}
 \caption{GOES 1-8\AA~lightcurves of each event. A pair of dotted vertical lines
represent the start and end times of the EIS raster and a single dashed line
marks the GOES derivative maximum for the event. The time at which the
spectrometer slit scans each footpoint is marked with a diamond on the light
curve.}
 \label{fig:goes}
\end{figure*}

Six events, varying in GOES class from B1.8 to C1.1, were selected from a period
between May and December 2007 (Figure \ref{fig:goes}). Two of the six events
labeled Event (b) (SOL2007-12-14T14:16:30) and Event (f)
(SOL2007-05-22T23:25:50) have been examined in greater detail; Event (b) by \citet{2009ApJ...699..968M} and 
\citet{2011ApJ...740...70M} --- on the subject of non-thermal line broadening, and Event (f) by \citet{2011A&A...526A...1D}. These events are observed to exhibit
footpoint EUV and HXR emission, chromospheric evaporation, and footpoint
electron density enhancements.

\begin{figure*}[t]
 \centering
 \subfloat{{(a)}\includegraphics[width=4cm]{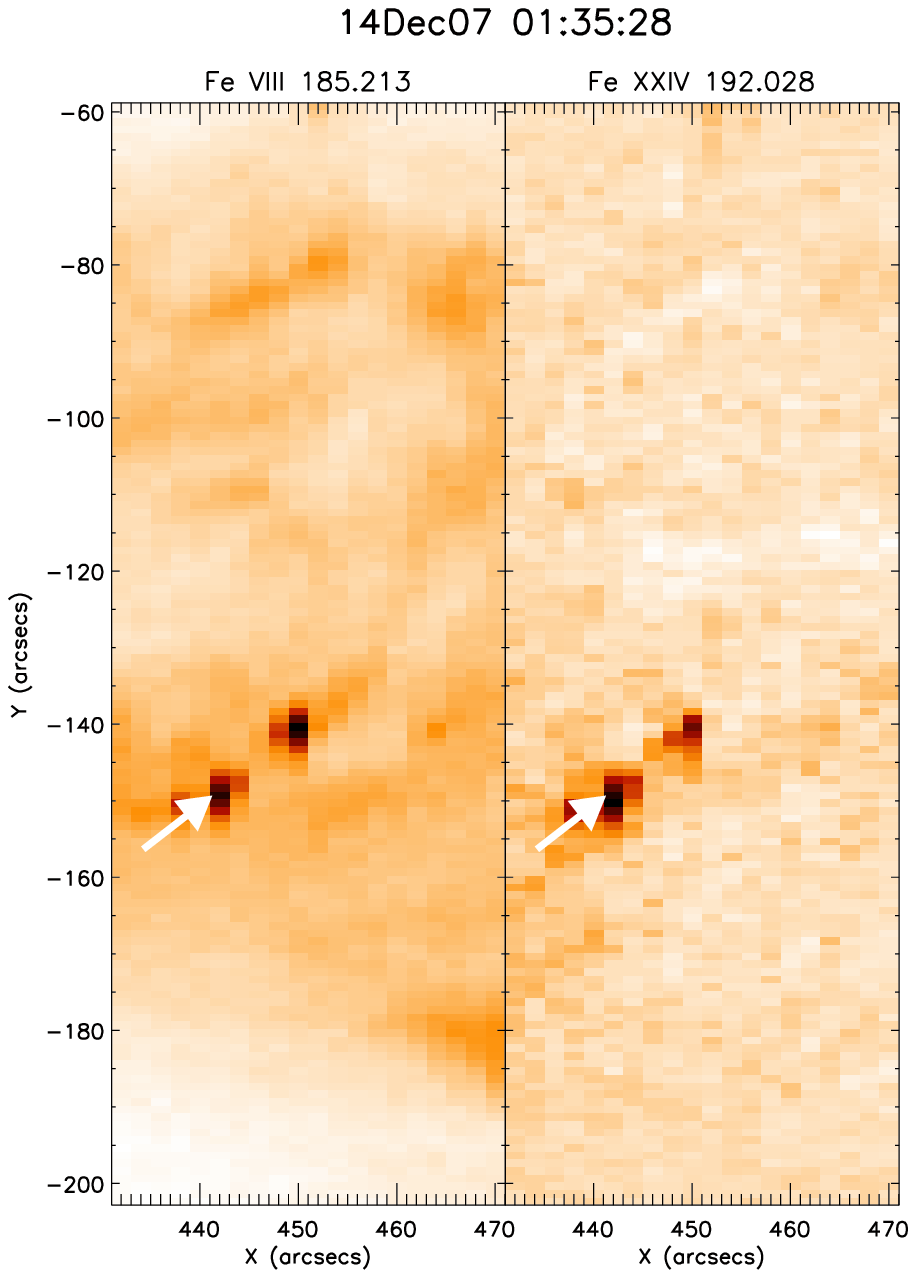}}
 \subfloat{{(b)}\includegraphics[width=4cm]{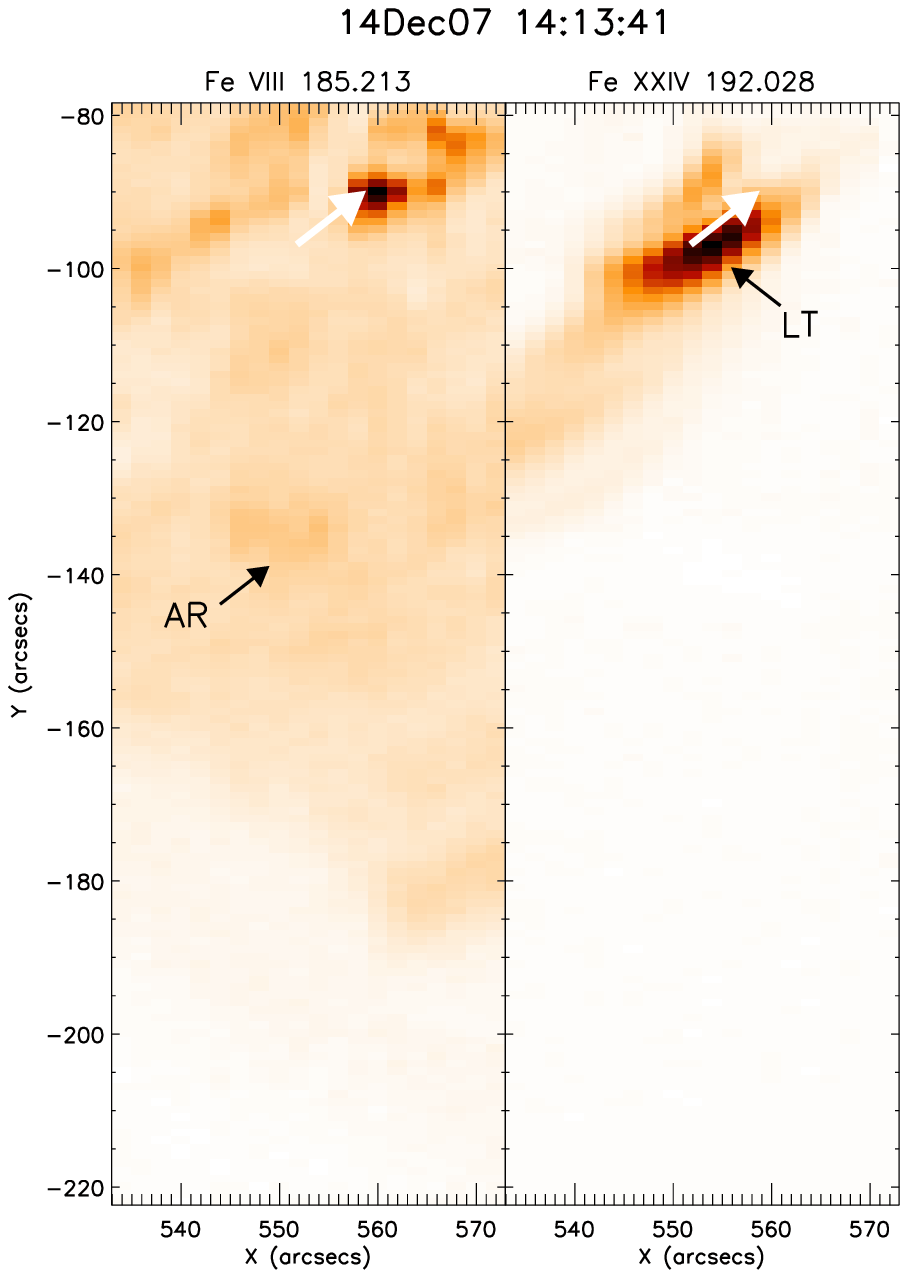}}
 \subfloat{{(c)}\includegraphics[width=4cm]{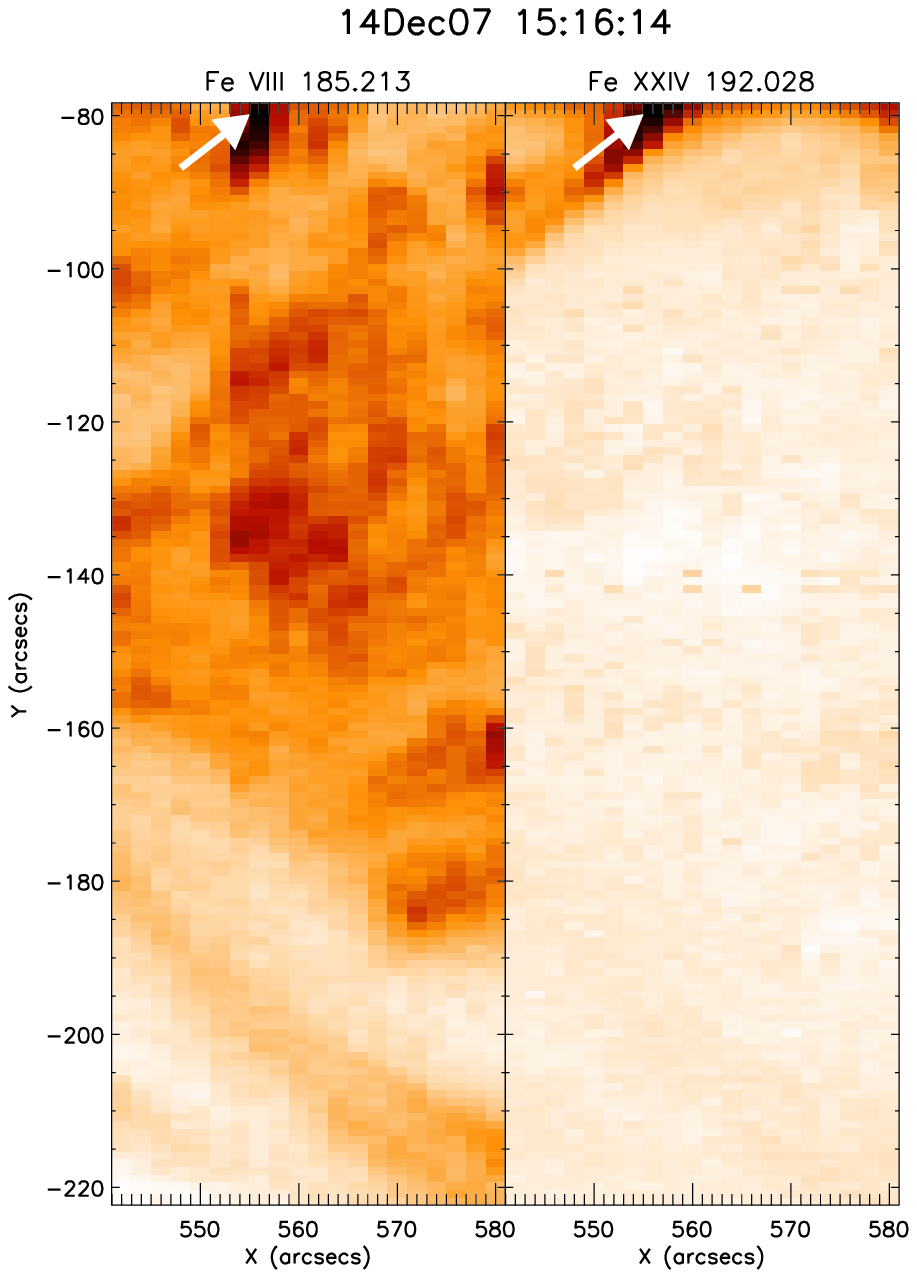}}

 \subfloat{{(d)}\includegraphics[width=4cm]{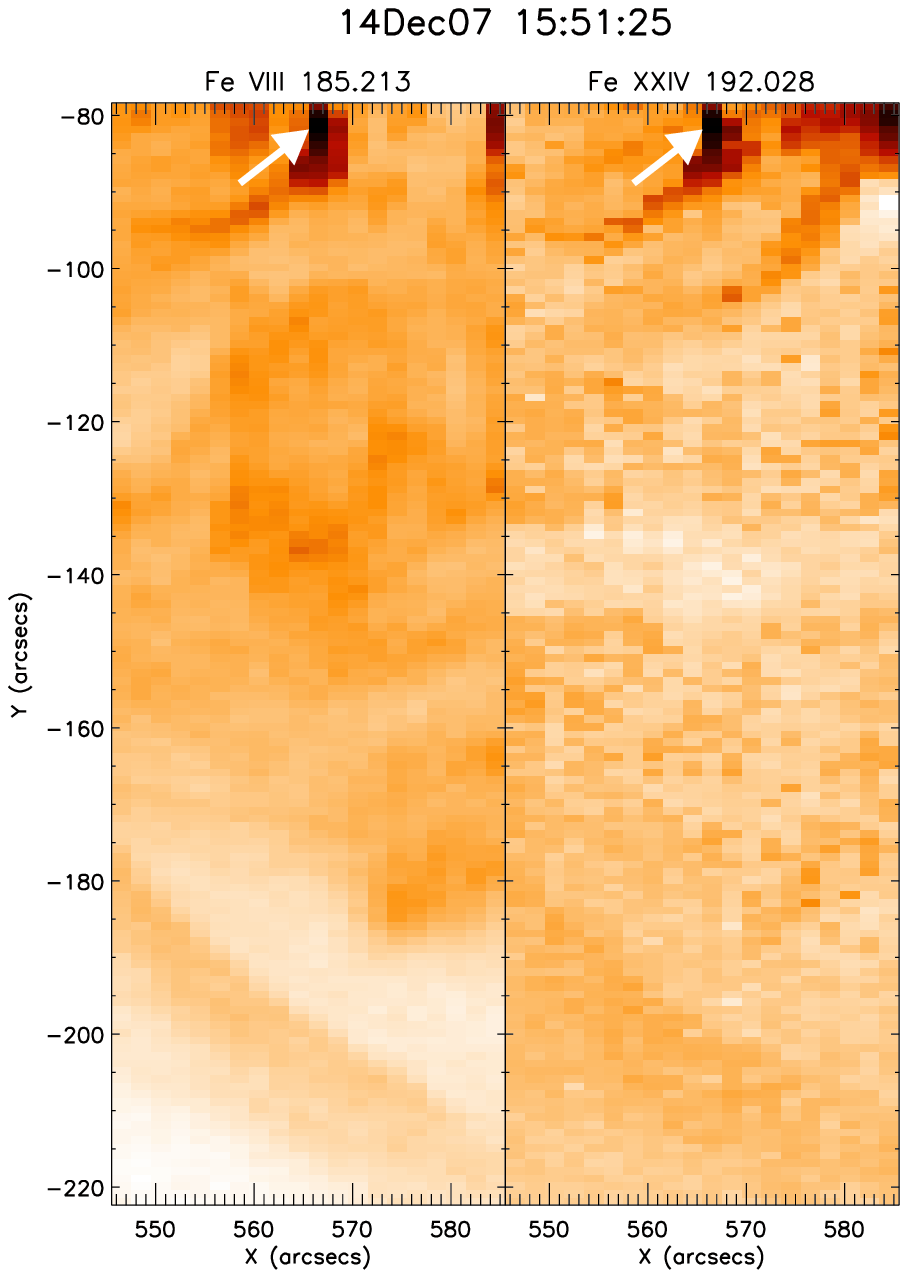}}
 \subfloat{{(e)}\includegraphics[width=4cm]{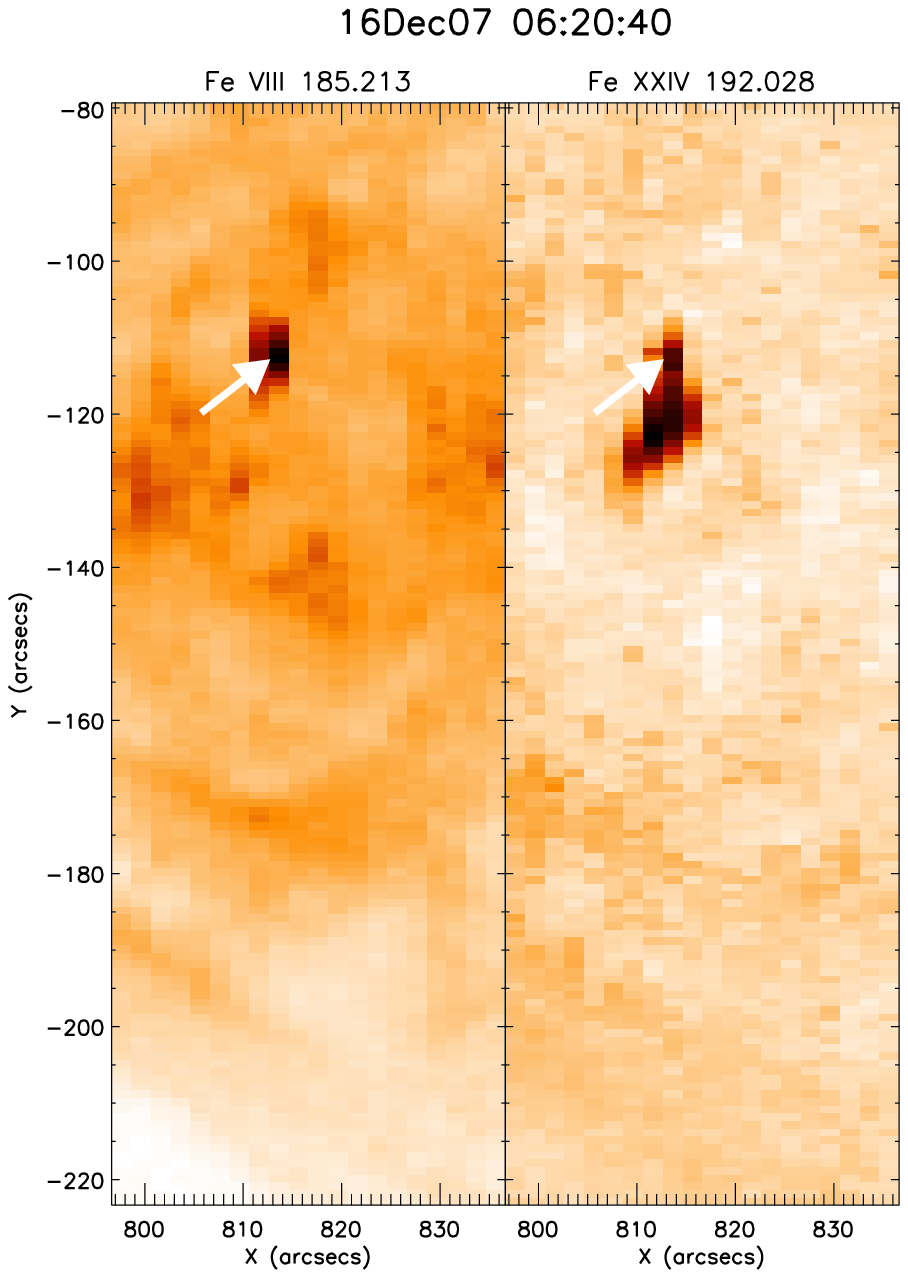}}
 \subfloat{{(f)}\includegraphics[width=4cm]{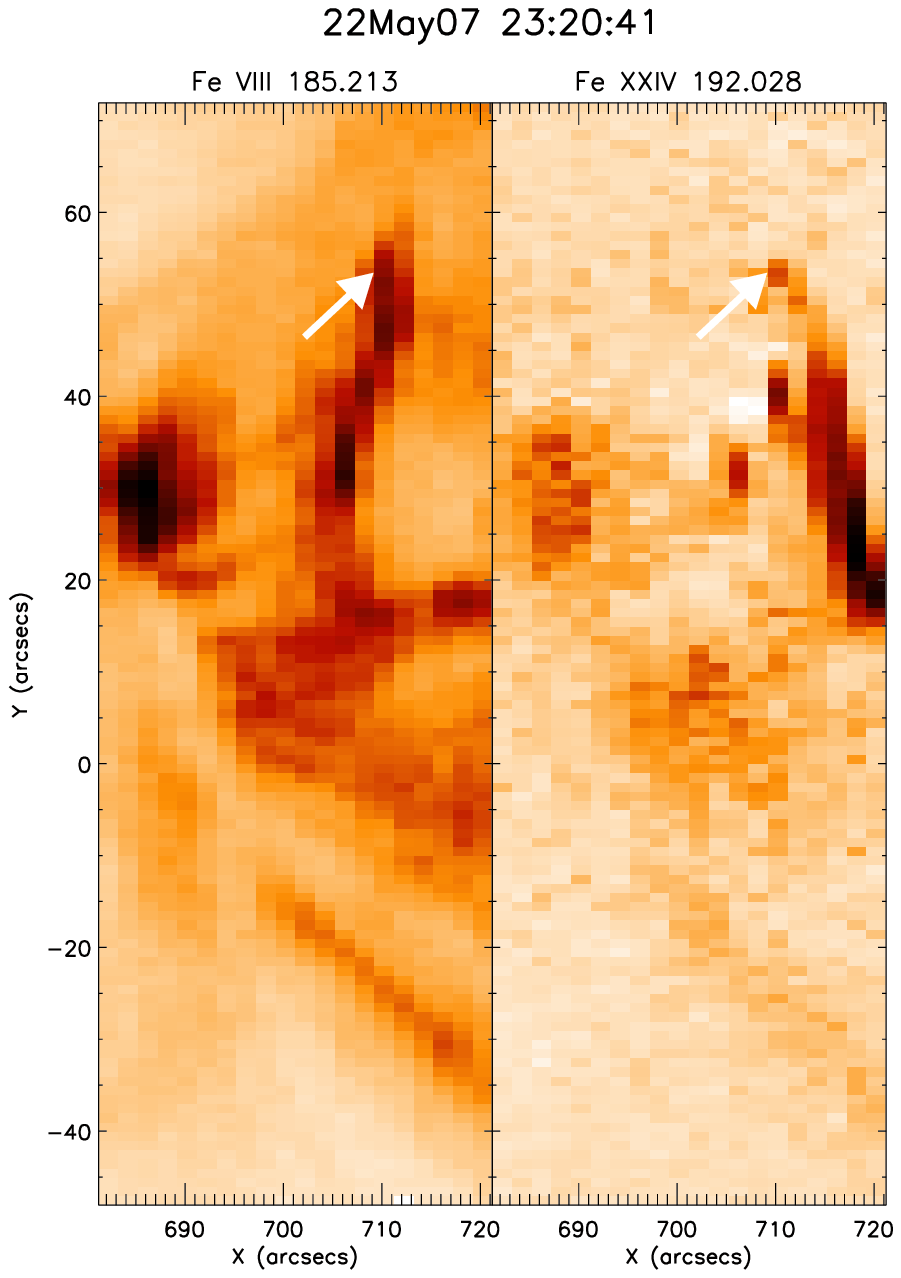}}
 \caption{EIS rasters for each event in Fe {\sc viii} at 500,000K and Fe {\sc
xxiv} at 16 MK showing the morphology at low and high temperatures. A white
arrow marks the footpoint positions chosen for EMD analysis in each raster.
Black arrows in panel (b) also highlight the regions used in determining a loop
top and active region EMD (see Section 6).}-
 \label{fig:hotcold}
\end{figure*}

Rasters shown in Figure \ref{fig:hotcold} illustrate the flare appearance at
500,000 K (Fe~{\sc viii}) and at 16 MK. (Fe~{\sc xxiv}).  From the work by
\citet{2009ApJ...699..968M} and \citet{2011A&A...532A..27G}, compact
brightenings in transition region lines can often be associated with RHESSI HXR
observations revealing the flare energy deposition site. Rasters have been
selected where compact, co-spatial Fe~{\sc viii} and Fe~{\sc xxiv} emission
rises dramatically compared to the background. Figure~\ref{fig:hotcold}a (Event
(a)) shows small bright sources appearing in Fe~{\sc viii} during the rise
phase. Compact, hot Fe~{\sc xxiv} emission is also present at this early stage
but becomes more significant later as evaporating hot material begins to fill
loop structures. Figure~\ref{fig:hotcold}b, (Event (b)) is sampled slightly
later in the impulsive phase and shows a hot flare loop forming next to the
footpoint.

These events are highly impulsive. GOES lightcurves for each event in Figure
\ref{fig:goes} show that the rise phase of most events lasts 2-4 minutes, with
the longest just under 10 minutes. Typically HXR observations are used to verify
that the EUV emission corresponds to the impulsive phase. However, RHESSI HXR
data were not consistently available, and so we systematically use the GOES
derivative as a proxy for the HXR emission \citep{1968ApJ...153L..59N}; a dashed
line on the lightcurves marks where this peaks and is used as a guide to select
EIS rasters during the impulsive phase. In each event at least one EIS raster
was found spanning part or all of the impulsive phase. Rasters have been chosen
as early as possible in the flare whilst still showing strong EUV enhancements,
and keeping the GOES derivative peak within the chosen raster limits (dotted
lines Figure \ref{fig:goes}). The flare evolves as the spectrometer slit scans
right to left over the window. The time at which the slit crosses a footpoint 
is marked by a 
diamond on the lightcurve.

Given this morphology throughout the EIS temperature range, plus supporting wide
field of view imaging 
from the X-Ray Telescope (XRT) onboard {\it Hinode} \citep{2007SoPh..243...63G}
and TRACE \citep{1999SoPh..187..229H} (omitted here for space constraints), we
are confident in identifying the flare footpoints, as marked with white arrows
in Figure~\ref{fig:hotcold}.

\section{DEM Technique}\label{sec:demtech}
The set of line observations, $I_\lambda$, and corresponding contribution
functions, $G_{\lambda}(T,n_e)$, are related to the differential emission measure of
the plasma via $I_\lambda=\int G_{\lambda}(T,n_e)\xi(T)\mathrm{d}T$ where the
DEM is defined as $\xi(T) =
n_e^2 (dS/dT)$. Obtaining $\xi(T)$ from $I_\lambda$ is an ``ill-posed'' inverse problem, with
uncertainties in the data resulting in the solution being non-unique. However, using physical constraints on the data can help recover a useful solution. There are a variety of techniques to recover the DEM
\citep[c.f.][]{1986SoPh..105..323F,1991AdSpR..11..281M} with a Markov chain
Monte Carlo (MCMC) approach being routinely adopted for spectral line data inversion
\citep{1998ApJ...503..450K}. In this paper we use a regularized inversion method
\citep{2012A&A...539A.146H}, which is able to produce solutions similar to the
MCMC approach but is computationally quicker and estimates both horizontal and
vertical errors on the DEM solution. Regularization adds a ``smoothness'' constraint to the DEM solution so that a stable inversion can be recovered, avoiding amplification of the uncertainties \citep{ti63}. However, this may be a strong assumption. For example, the DEM of a loop in conductive equilibrium
theoretically has a discontinuous, high temperature cut-off.
The minimum of the EM loci curves (the curves representing the isothermal
emission
in each line) are used as an initial guess DEM solution. A multi-thermal DEM
solution must be below these EM loci curves (since the isothermal solution gives
the maximum possible emission at that temperature) and the regularized solutions
achieve this (see Figure \ref{fig:footdem}). The approach iterates until a
positive DEM is found and also minimizes the chi-squared between the measured
and regularized line intensities. For a full explanation of this approach on EIS
line data compared to the MCMC method see \citet{2012A&A...539A.146H}.

\subsection{DEM Inputs and Uncertainties}\label{sec:errors}

Determination of the DEM requires three inputs; line intensities, intensity
error estimates, and the line contribution functions, here calculated using
CHIANTI v7.0. These are calculated using coronal abundances
\citep{1992ApJS...81..387F}, CHIANTI's own ionization equilibrium file
\citep{2012ApJ...744...99L} and a constant density of $10^{11} {\rm cm^{-3}}$
suitable for flare footpoints  \citep[see e.g. ][]{2010ApJ...719..213W,
2011A&A...532A..27G, 2011ApJ...740...70M}. The DEM shape should not be strongly
affected by density variations, as the density-sensitive lines have been removed
from the analysis. We assume a 20\% systematic error across all intensity
measurements to account for the absolute calibration uncertainty between lines
(P. Young, private communication) and this is added to the fitting error.

It is difficult to determine the correct elemental abundances for use in flare
analysis \citep{1994ApJ...423..516A, 2004ApJ...609..439F}. In the standard
model, footpoint material originates low in the chromosphere before being heated
to coronal temperatures, yet this material rises and mixes with existing coronal
material in loops. The choice of abundances used to interpret flare spectra is
therefore not straightforward. Furthermore, low FIP (first ionization potential)
elements are found to be enhanced in coronal material compared to the
photosphere while high FIP elements are unchanged. Only O {\sc v} and O {\sc vi}
in our analysis are high FIP.  The choice of ionization equilibrium is similarly
uncertain. We will investigate the effects of varying abundance and ionization
equilibrium in Section~\ref{sec:abun_ion}

\subsection{Assumptions}\label{sec:assumptions}

\begin{figure*}[h!]
 \centering
 \subfloat{{(a)}\includegraphics[width=5.5cm]{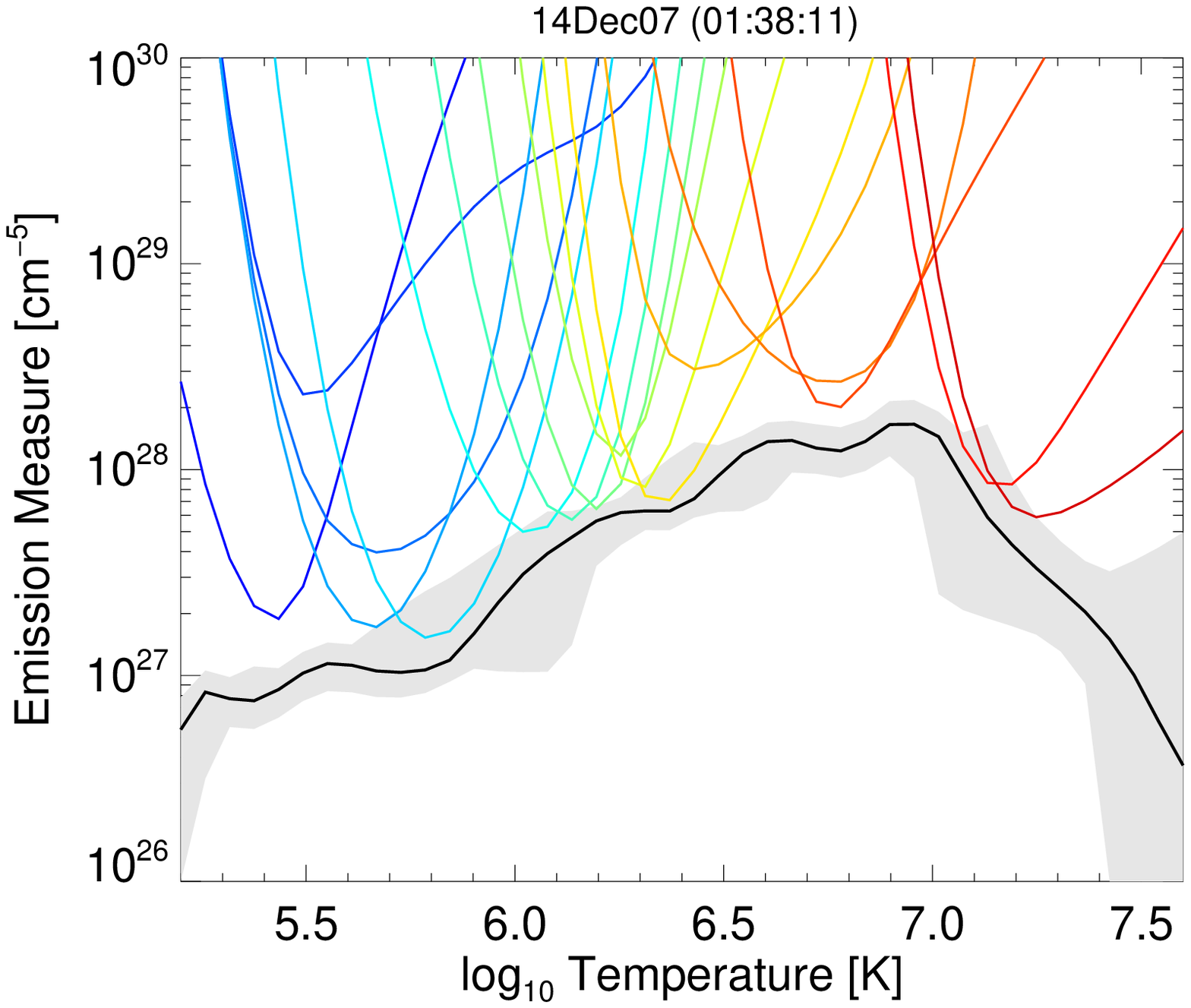}}
 \subfloat{{(b)}\includegraphics[width=5.5cm]{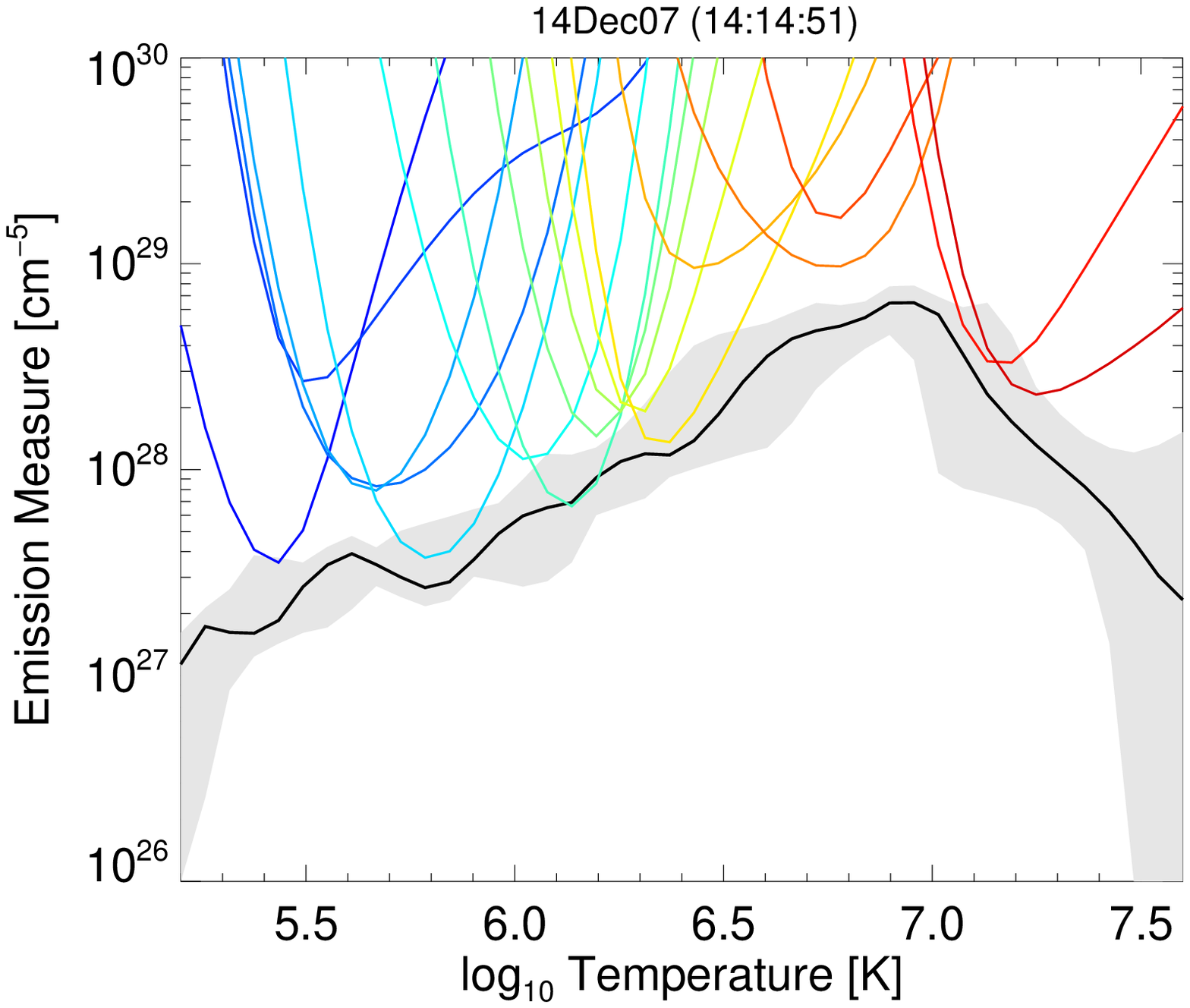}}
 \subfloat{{(c)}\includegraphics[width=5.5cm]{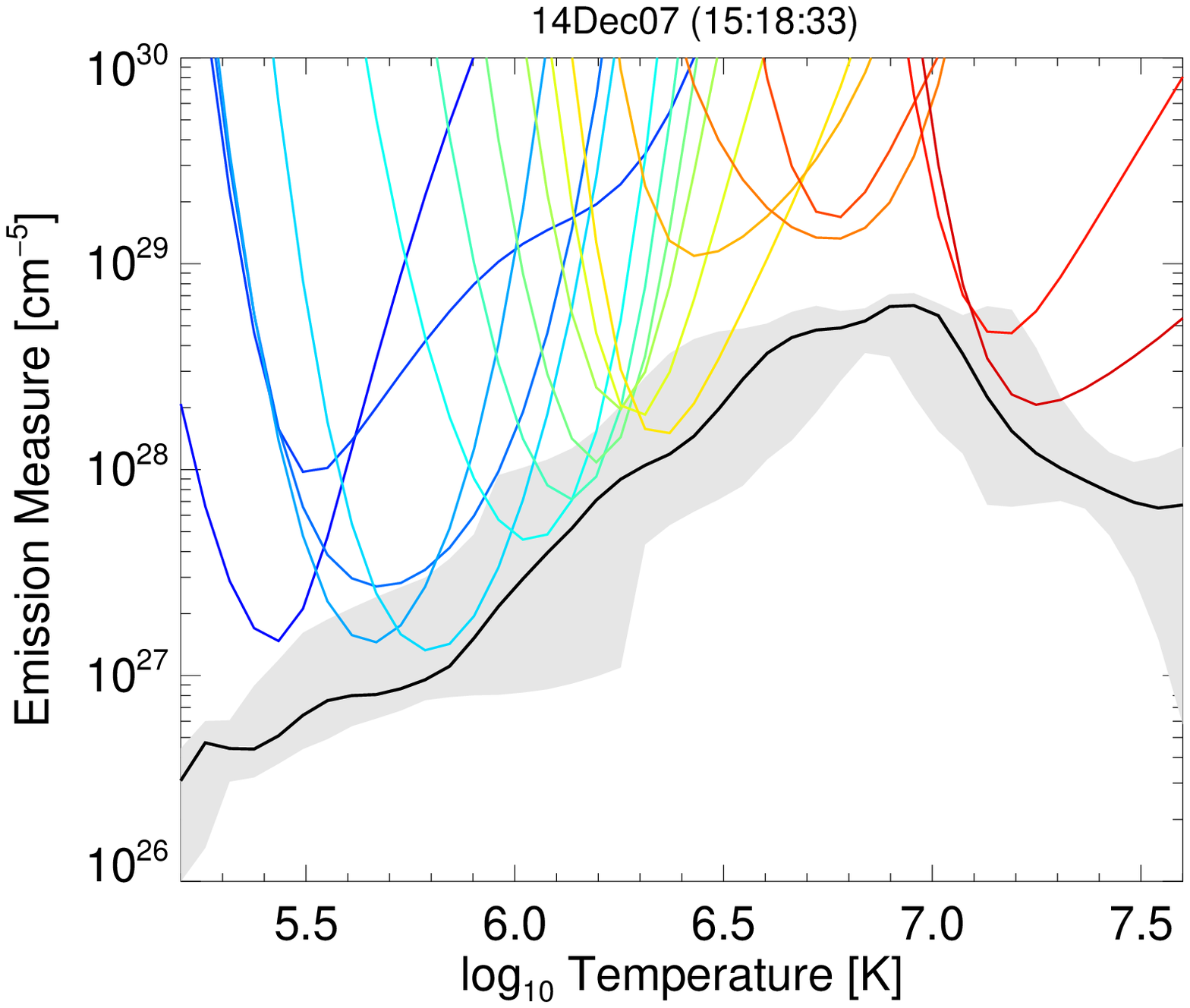}}

 \subfloat{{(d)}\includegraphics[width=5.5cm]{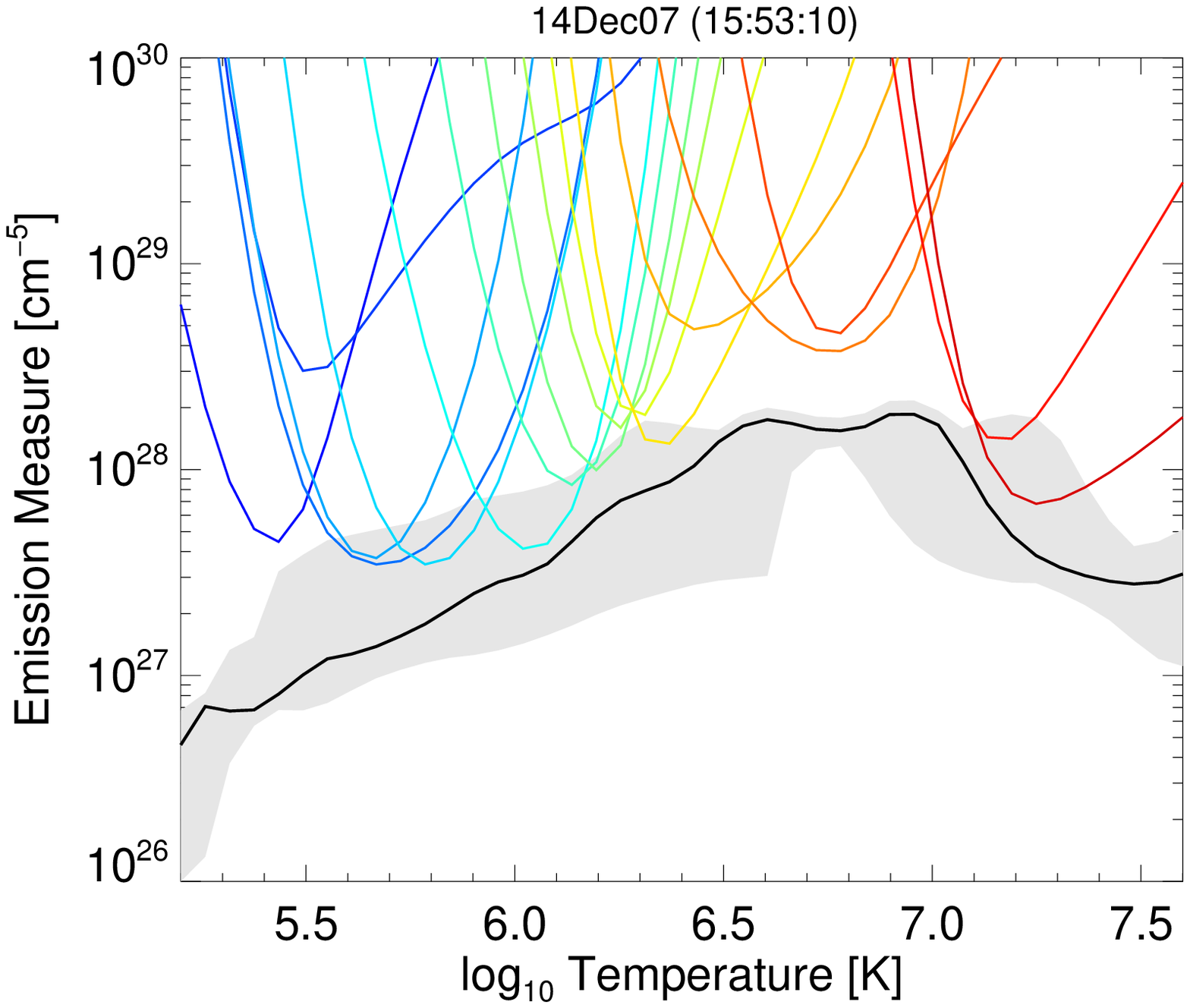}}
 \subfloat{{(e)}\includegraphics[width=5.5cm]{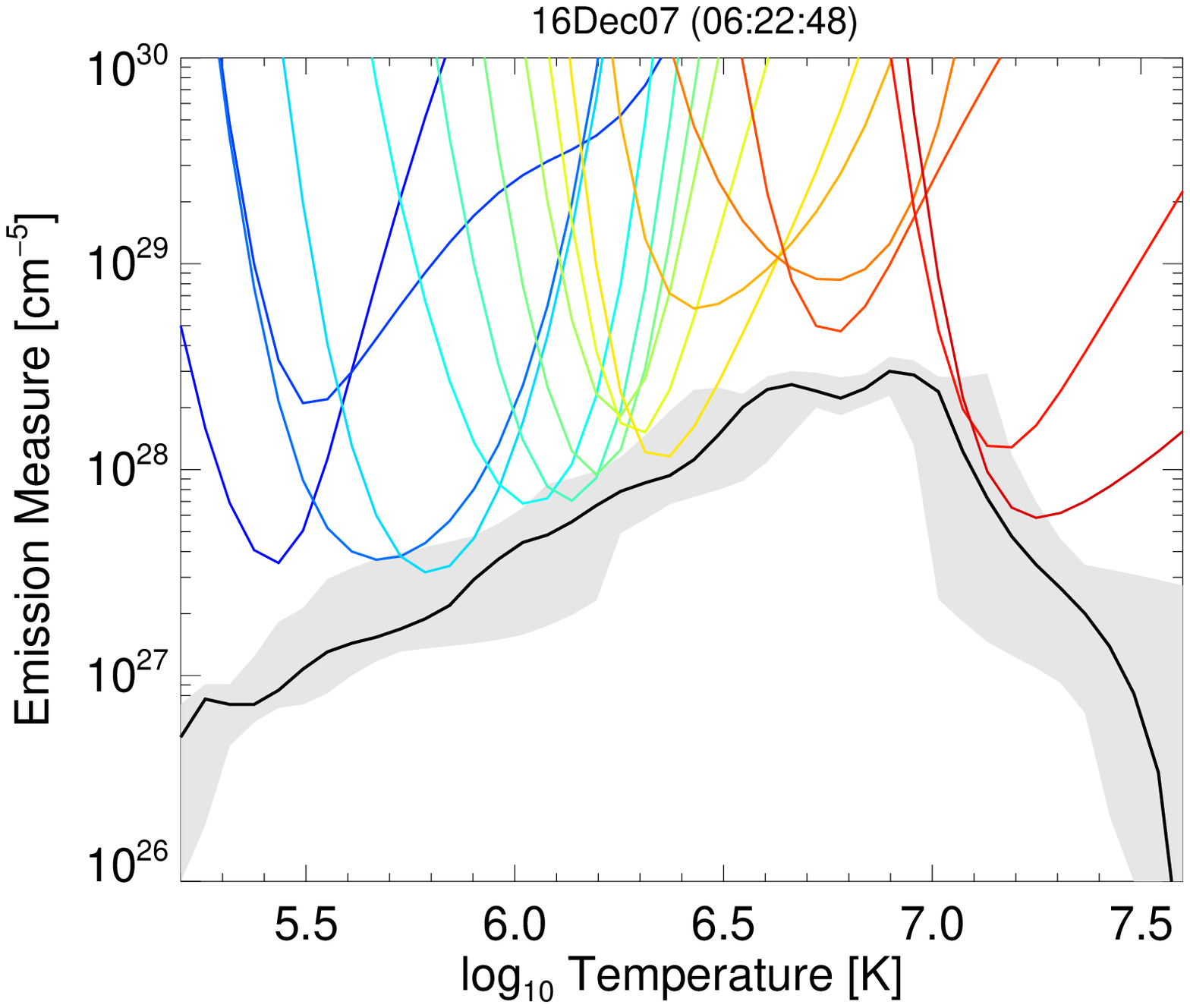}}
 \subfloat{{(f)}\includegraphics[width=5.5cm]{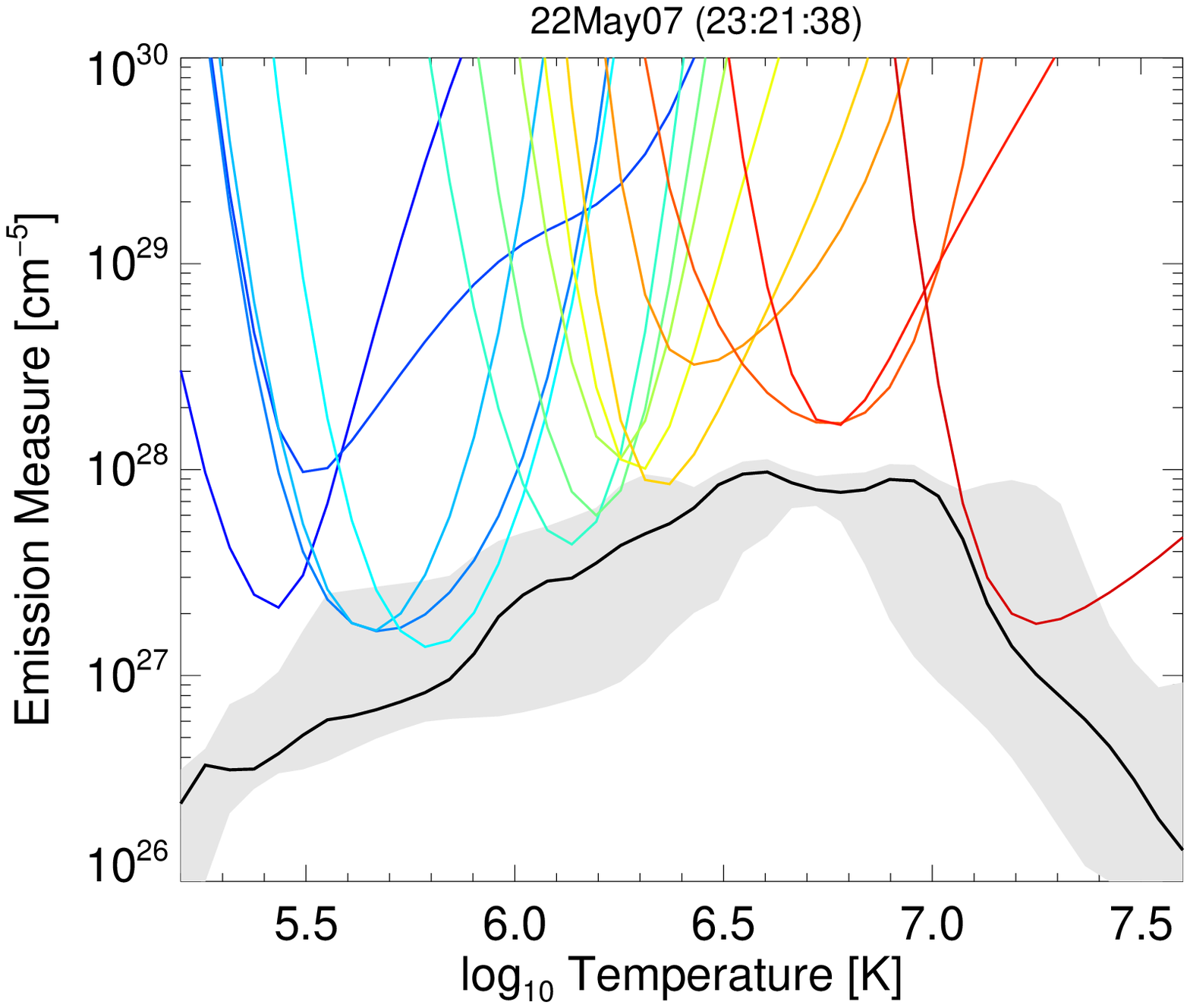}}
 \caption{DEMs shown in black for each event in emission measure units cm$^{-5}$
and the uncertainty limits of the solution by a shaded grey area. The colored
curves show measured line intensity divided by the contribution function
indicating the maximum possible emission, i.e. the EM loci curves.}
 \label{fig:footdem}
\end{figure*}

Before interpreting spectra with CHIANTI one must be aware of the built-in
assumptions. The $G(T)$ functions are calculated for an optically thin plasma in
thermal and ionization equilibrium. Clearly from RHESSI HXR observations flare
footpoint electron spectra have an inherently non-Maxwellian component of the
electron distribution, although this can be small compared to the total energy
of the distribution; see \citet{2011ApJ...739...96K}. In addition the footpoint
is a location of intense heating, so it is possible to assume that the plasma is
out of thermal and ionization equilibrium. However, the high density of the
emitting region may provide a high enough electron collision rate for the
footpoint plasma to be close to equilibrium. For a plasma at $10^{6}$ K and $n_e
= 10^{11} {\rm~cm^{-3}}$ the electron-electron collision timescale is $\tau_{ee}
= 1.33\times 10^{-4}~s$, which is probably much shorter than the flare heating
timescale, so the core electrons of the distribution can rapidly reach thermal 
equilibrium.

Calculation of non-equilibrium ionization (NEI) states by
\citet{2009A&A...502..409B} shows that during heating the population of a given
ion peaks at a higher temperature than that at which it would peak in
equilibrium. This could lead to systematically lower temperatures being
inferred. Again the effect is more pronounced at lower densities as ionization
and recombination processes are driven by electron collisions.
\citet{2009A&A...502..409B} finds that at densities of $10^{10} {\rm~cm^{-3}}$
the peaks can shift by up to $\rm~{log~T} = 0.3~K$ but by $n_e = 10^{12}
{\rm~cm^{-3}}$ the difference is negligible.

Finally we assume that the plasma is optically thin. This may at first sight be
a more problematic assumption given the high densities and footpoint emission
originating from deeper regions of the atmosphere, which may be optically thick.
 We estimate the opacity for Fe {\sc viii}, Fe {\sc x} and Fe {\sc xvi} from the
method in \citet{2002A&A...390..219B} and \citet{2011ApJ...740...70M}. The
optical depth at line centre, $\tau_{0}$ can be expressed as

\begin{equation}
\tau_{0} = 1.16 \times 10^{-14} \lambda f_{ij} \sqrt{M\over T} {n_{ion}\over
n_{el}} {n_{el}\over n_{H}} {n_{H}\over N_{e}} N_{e} h
\label{opacity}
\end{equation}

where $\lambda$ is the line wavelength, $f_{ij}$ the oscillator strength of the
transition, and $M$ the mass of the ion. ${n_{ion}/n_{el}}$ and ${n_{H}/N_{e}}$
are taken from the default CHIANTI ionization equilibrium file and coronal abundances.
Taking a path length of 1\arcsec\ and $n_H = 5 \times 10^{9} {\rm~cm^{-3}}$
we find $\tau_0 = 0.527, 0.06 {\rm ~and~} 0.04$ respectively for Fe {\sc viii}, Fe {\sc x} and Fe {\sc xvi}.
As expected emission lines at higher temperatures are largely unaffected by the plasma opacity,
yet the cooler Fe {\sc viii} line may be influenced. An optical depth of $\tau_0 =
0.527$ corresponds to a drop in transmission of $\sim 40\%$ --- compared to the line
uncertainty of at least 20\% this could be significant. However, if photospheric abundances are used this
drops to $\sim 10\%$, demonstrating that a careful analysis of flare abundances is required in future. To fully understand the effects of non-equilibrium and
radiative transfer in a flare atmosphere is a serious undertaking which lies
beyond the scope of this paper. Our results are the first of their kind using
the best spectroscopy and atomic calculations available, but we must be aware of
these possible shortcomings.

\section{Flare EMDs}

EMDs derived for the six different events using the above methods are shown in
Figure \ref{fig:footdem}. The figures show the emission measure distribution
$EM(T)$ in units of cm$^{-5}$ found by integrating the DEM over a fixed
logarithmic temperature interval. This returns an EMD of the same form as those
discussed in the introduction. In all events the EMDs in black lines are bounded
by the colored EM loci curves, confirming that the regularized solutions are
below the expected maximum emission.

A shaded region outlines the extent of the EMD uncertainty in temperature and EM
space. The true solution lies within this boundary. Within the plotted
temperature range the uncertainties in temperature are mostly within an order of
magnitude or less in EM. At temperatures above $10^{7}$~K the solutions have a
large uncertainty due to the broad $G(T)$ response in Fe {\sc xxiii} and Fe {\sc
xxiv}. Also these emission lines are fainter at footpoints so have larger
fitting errors. Unsaturated, soft X-Ray observations from {\it Hinode}/XRT would
have helped constrain this part of the EMD but were unavailable. Extending the
temperature range beyond the limits of the $G(T)$ functions significantly
spreads out the errors at the temperature limits. The regularization is unable
to find solutions where $G(T)$ is undefined, hence it is therefore not possible
to make physical conclusions beyond these limits.

All of the footpoint EMDs share a strikingly similar profile: increasing with an
almost constant gradient of $EM(T) \sim T$ to a peak around $\log~T = 6.9$ then
falling off quickly at higher temperatures. The peak temperature suggests a
significant presence of plasma at 8 MK in the flare footpoints. Peak emission
measures vary between $10^{28}$ and $10^{29}$ cm$^{-5}$ but do not appear
strongly related to the GOES class, which is perhaps expected for a small sample
of lower energy events sampled at slightly different times in their evolution
(see Table \ref{tab:events}).

\begin{figure}[h!]
 \centering
 \includegraphics[width=8.2cm]{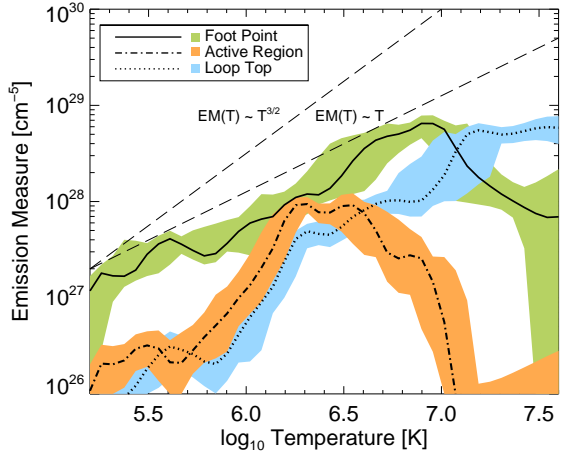}
 \caption{A comparison of footpoint, loop, and active region EMDs in Event (b).
The purpose of this is to check the regularization response to different
temperature distributions. Gradients of $EM(T) \sim T$ and $EM(T) \sim
T^{\frac{3}{2}}$ are added in dashed lines.}
 \label{fig:demcomp}
\end{figure}

We measure a gradient of $EM(T) \sim T^{0.97\pm{0.27}}$ between $\log T = 5.5 -
6.9$ in Event (b), where the uncertainty is estimated by the maximum and minimum
gradients allowed within the EMD error region. This footpoint EMD is plotted in
Figure \ref{fig:demcomp} (green region) against a line of gradient $EM(T) \sim
{T}$ resembling our event, and $EM(T) \sim T^{\frac{3}{2}}$, a commonly observed
gradient for a transition region/low corona atmosphere.

To verify that the similarity of the footpoint EMDs is not an artefact of the
regularization method used, we have calculated EMDs from both active region (AR)
and flare loop top (LT) plasma. Again using Event (b), EMDs from AR and LT
locations are shown in Figure \ref{fig:demcomp} in the orange and blue regions
respectively, with the footpoint shown in green. The AR and LT locations on the
raster are marked for reference by black arrows on Figure \ref{fig:hotcold}b.
Their EMDs are noticeably different from the footpoint EMD, showing that the
regularization responds well to different plasma temperature distributions. The
active region EMD peaks between $\log T = 6.3 - 6.5$ with a lower emission
measure and steep high temperature cut-off. The gradient in the AR and LT EMDs
are steep ($> T^{\frac{3}{2}}$) and remarkably similar up to $\log T = 6.3$
where the loop top EMD becomes shallower and continues rising to over 10 MK ---
the point at which the EMD becomes poorly constrained.

Emission in the footpoint between temperatures of $\log T = 5.2 - 6.2$ is greater by up to an
order of magnitude than in the active region or loop top. From the conventional 
understanding of flares it is likely this is chromospheric plasma in the process 
of being heated to flare temperatures. The break in similarity between the LT
and AR gradients at $\log T = 6.3$ suggests that only plasma above this temperature
is being evaporated into the flare loop, agreeing with the results in \citet{2009ApJ...699..968M}
where only footpoint plasma above $\log T = 6.2$ exhibits evaporative upflows. A
full flow velocity analysis will be the subject of future work, but it is
reassuring to see that such arguments can be made from the results and that the
behaviour of the EMDs varies in reasonable ways across the event.

\begin{table*}[h!]
\caption{Event EMD parameters sorted by GOES Class.}\label{tab:events}
 \begin{center}
 \begin{tabular}{ccccc}
\tableline\tableline

Flare ID & GOES Class & Event & Peak Temp & Peak EM \\
 &  &  & $log_{10}~T$ (K) & $\times 10^{28} {\rm cm^{-5}}$ \\
\tableline

SOL2007-12-16T06:22:40 & B1.8 & (e) & 6.9 & 4.0 \\
SOL2007-05-22T23:25:50 & B2.7 & (f) & 6.9 & 1.0 \\
SOL2007-12-14T15:22:00 & B8.8 & (c) & 6.9 & 8.0 \\
SOL2007-12-14T15:54:15 & B8.8 & (d) & 6.9 & 2.0 \\
SOL2007-12-14T01:39:20 & B9.6 & (a) & 6.9 & 2.0 \\
SOL2007-12-14T14:16:30 & C1.1 & (b) & 6.9 & 6.0 \\

\tableline
\end{tabular}
\end{center}
\end{table*}

\subsection{Varying the abundance and ionization
equilibrium}\label{sec:abun_ion}
\begin{figure}[h!]
 \centering
 \includegraphics[width=8.2cm]{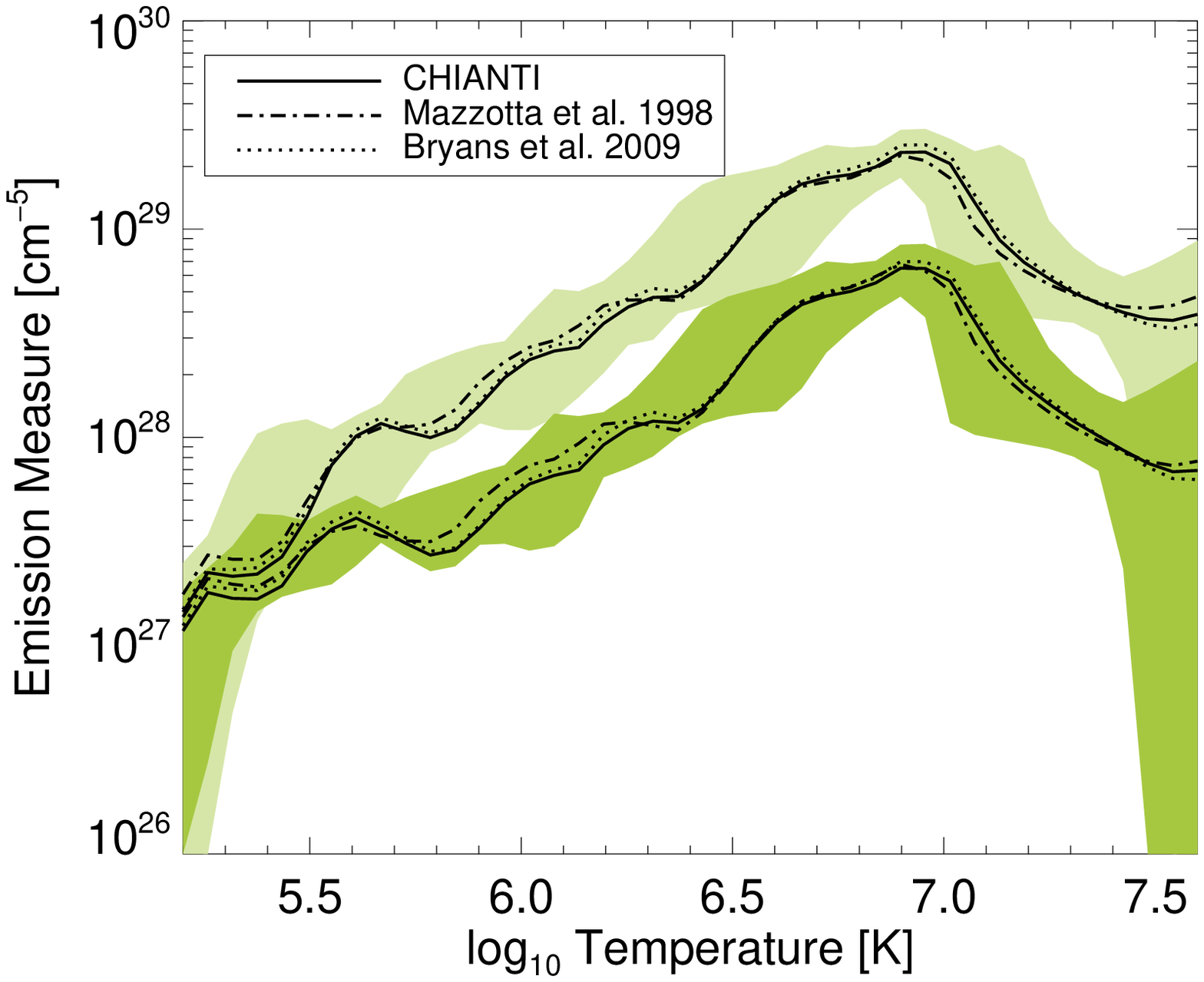}
 \caption{Footpoint EMD of Event (b) using photospheric and coronal abundances
and a variety of ionization equilibrium theories \citep{2012ApJ...744...99L,
1998A&AS..133..403M, 2009ApJ...691.1540B}. EMDs are shown by the black lines
within a shaded error boundary. Curves within the lighter shaded area are from
photospheric abundances and the darker from coronal abundances. For each
abundance file three ionization equilibrium theories are tested and plotted in
different line styles.}
 \label{fig:demparam}
\end{figure}

As discussed in Section~\ref{sec:errors}, the appropriate values of abundance in
the chromospheric plasma, and of the ionization equilibrium, are not known,
therefore we repeat the analysis using various ionization files with
photospheric and coronal abundances (\citet{1998SSRv...85..161G} and
\citet{1992ApJS...81..387F} respectively). EMD curves for the footpoint in Event
(b) are shown in Figure \ref{fig:demparam} using photospheric (light shading)
and coronal (dark shading) abundances. The EMD is around a factor of 4-5 larger
using photospheric abundances but varies very little in shape; only at very low
temperatures where oxygen is dominant is there any deviation in gradient.
Changing the abundances therefore only alters the result significantly by a
constant factor in EM, i.e. the gradient remains $EM(T) \sim T$ between $\log T
\sim 5.5 - 6.9$.

Looking again at Figure \ref{fig:demparam} changing the ionization equilibrium
parameters (solid, dashed, and dotted line styles) also has very little effect
on the EMD. Given the large number of lines and small variations these
parameters produce on the contribution functions, the final EMD is relatively
insensitive, especially when considering how much larger the intensity
uncertainties are. Any differences due to different ionization equilibria all
lie within the error boundaries plotted.

\subsection{Column emission measures}\label{sec:column}

Estimating the emission measure of emitting plasma can also be approached
through the use of density-sensitive line ratios. \citet{2011ApJ...740...70M}
used five line ratios available in these rasters (Mg {\sc vii}, Si {\sc x}, Fe
{\sc xii}, Fe {\sc xiii}, and Fe {\sc xiv}) to estimate the footpoint electron
density at various temperatures in Event (b), and from these and the observed intensities, calculate the respective column depths of the emitting material. Here we use these measurements of column depth for an alternate estimate of the column emission measure at a range of temperatures.
 
The intensity of a given emission line, $I$, integrated over the line of sight column depth, $S$, can be expressed as:
\begin{equation}
4 \pi I = 0.83 \int G(T, n_{e}) n_{e}^{2} dS.
\label{col_depth_one}
\end{equation}

\noindent                                                                                                                                                                                                                                             
By assuming the electron density, where $n_e$ is obtained from independent density diagnostic pairs, is constant across each pixel, and calculating the line contribution function, $G(T,n_{e})$, at the measured electron density, the column depth is derived (see \citet{2011ApJ...740...70M} for further details). Since the column emission measure $EM_{col}$ is defined by
$EM_{col}=\int~n_{e}^{2} dS$, we have combined the electron density and column
depth measurements to estimate $EM_{col}$ for each diagnostic line pair at the
footpoint in Event (b). 

\begin{figure}[!t]
\begin{center}
\includegraphics[width=9.1cm,angle=90]{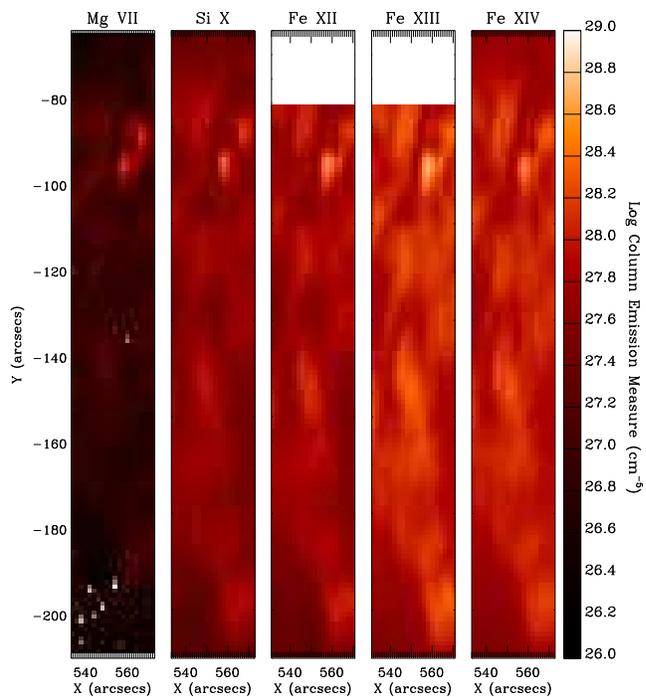}
\caption{\small{The reconstructed column emission measures in Event (b), found
by combining the measured electron densities and column depths for the Mg VII,
Si X, Fe XII, Fe XIII, and Fe XIV lines. The footpoint can be seen at (-90\arcsec\, 560\arcsec). \label{fig:column_emiss}}}
\end{center}
\end{figure}

The 5 panels in Figure~\ref{fig:column_emiss} show maps
of column emission measure for Mg {\sc vii}, Si {\sc x}, Fe {\sc xii}, Fe {\sc
xiii}, and Fe {\sc xiv}. In each of the five rasters, higher column emission
measures were found at the footpoint locations compared to the surrounding
active region, even in the cooler Mg {\sc vii} and Si {\sc x} lines. The footpoint column emission measures returned from Figure~\ref{fig:column_emiss} are between $10^{28}-10^{29} \rm cm^{-5}$, in agreement with the regularized inversion method. Uncertainties in both the density estimates and
regularized EMDs make it difficult to comment on deviations between the
techniques of less than an order of magnitude. However, as the observed
deviations are not larger than this, the column 
emission measures found using line diagnostics do help to reinforce the emission
measures obtained via the inversion method.

\section{Discussion}\label{sec:discussion}
Different theoretical models of energy inputs and losses, during flares and in
the quiet Sun, produce EMDs with different slopes,  and the slope may provide a
diagnostic of the energy balance of the emitting plasma. During a flare the
energy balance equation can be very complex, so it is normal to make certain
simplifying assumptions - for example, that the temperature and ionization
fraction of the plasma has reached a steady state, that the plasma is
gravitationally stratified, and that the emitting structure has constant
cross-section. One of the earliest treatments by \cite{1973SoPh...33..341S}
investigated a flare atmosphere split into two layers. In the upper (high
temperature) layer, flare energy deposition occurred, and was balanced by
conductive losses to the lower layer. Conduction was assumed to dominate
radiative output from this layer. In the lower layer, which did not receive any
direct flare heating (e.g. by electrons), the balance was between conductive
input from above and radiative losses.

Interpreting Figure 5 in \cite{1973SoPh...33..341S}, the resulting EMD gradient
in a $\log~EM - \log~T$ plot using a constant pressure assumption was  $\sim
1.2$ over a range of a few $\times 10^5$ to $10^7$K. (It is $\sim 2.2$ using a
constant density assumption). They argue that constant pressure was a valid
assumption when dealing with the narrow (compared to pressure scale height)
high-temperature transition between the flaring chromosphere and corona.

A similar analysis in \cite{1982ApJ...258..835W}, looking only at the layer
where conduction and radiation balance,  demonstrated a $\log EM-\log~T $ slope
of 1, as we observed. The difference between this result and that of
\cite{1973SoPh...33..341S} may lie in the form of the radiative loss function
adopted.

The agreement between the slopes found by \cite{1973SoPh...33..341S},
\cite{1982ApJ...258..835W}, and those derived from our observations,  is
intriguing. Perhaps it suggests that in our events energy injected at the
footpoints is localized to the very top of the flare chromosphere, in a region
at a temperature of $\log T  \gtrsim 6.9$, with the temperature structure
beneath determined primarily by conduction and radiation. In other words, any
direct flare energy input in the region $\log T \sim 5.5 - 6.9$ is negligible in
magnitude compared to other energy loss or gain terms.
\citet{2012ApJ...754...54B} has inferred similar behaviour in a C6.6 class
flare. The early appearance of coronal Fe {\sc xix} emission and late rise of
transition region lines (O {\sc V}, Si {\sc xii} and He~{\sc i}) was interpreted
as evidence for transition region plasma being heated by thermal conduction from
directly heated coronal plasma.

However the agreement between model and observation may also be coincidental, as
there are other assumptions in the analyses discussed. For example, it is
assumed that conduction is determined by classical Spitzer conductivity, but the
strong temperature gradients implied by the small vertical extent of the flare
transition region mean that non-local or saturated flux effects may be important
\citep[e.g.][]{2009A&A...498..891B}.  Mechanical energy loss and enthalpy 
flux due to plasma flows (e.g. evaporation) are also neglected. These loss terms
may not be important everywhere in the EMD temperature range;
\cite{2009ApJ...699..968M}  show that in Event (b) high speed evaporative
upflows are only present above $\sim 2$~MK. \cite{1978ApJ...224.1017U} show that
the effect of evaporation tends to be to flatten the slope of the EMD.  Further
work is needed to understand the effects of relaxing these assumptions, as well
as on exploring other descriptions for the flare direct energy input and the
radiative loss function.

\section{Conclusions}
We have obtained the first emission measure distribution of the plasma at a
flare footpoint using data from {\it Hinode}/EIS in conjunction with a
regularized inversion method. The spectral imaging capabilities of EIS allows us
to separate the footpoint EUV spectra -- therefore EMD -- from loop structures;
this ambiguity has been a drawback in many previous studies. The resulting
footpoint EMDs can be characterized by an emission measure gradient of $EM(T)
\sim T$ between $\log T \sim 5.5 - 6.9$ that falls away at higher temperatures,
and peak emission measures on the order of $10^{28} - 10^{29} \rm~cm^{-5}$. The
absolute value of emission measure in the EMDs is further confirmed by the use
of density-sensitive line ratios to estimate footpoint column emission measures.
In previous theoretical work the EMD gradient is found to be sensitive to the
energy transfer methods in a heated atmosphere. Our EMD profiles are in rough
agreement with a flaring mechanism depositing energy at the top of the flare 
chromosphere heated to $\log T \gtrsim 6.9$. Deeper layers then radiate the
conductive flux received from the hot layer above.

Obtaining reliable measurements of footpoint EMDs is not only of theoretical
interest, as they can be used to better estimate synthetic line intensities in
flares. This is useful for identifying blends and lines in other instruments
such as the SDO/EVE spectrometer, or in calibrating instrumental responses for
future instruments (see the forthcoming IRIS mission).

The next step in understanding the heating of footpoint plasma is naturally to
use RHESSI HXR observations to estimate the energy deposited in the
chromosphere. How does the EMD depend on the HXR spectra of the event, are they
as consistent as the EMD themselves? Combined with the temperature response in
EUV lines a fuller picture would be available to compare various flare heating
models.

\acknowledgments

The authors thank Hugh Hudson and Peter Young for their insightful comments and
discussions. We are also grateful to an anonymous referee whose comments led to new insight in interpreting the EMDs. This work was supported by STFC grant ST/I001808/1, by Leverhulme grant F00-179A
and by EC-funded FP7 project HESPE (FP7-2010-SPACE-1-263086). DRG acknowledges
support from an STFC-funded PhD studentship and ROM is grateful to the
Leverhulme Trust for financial support from grant F/00203/X, and to NASA for
LWS/TR\&T grant NNX11AQ53G. LF \& ROM are grateful to ISSI being where the early stages of this work were explored. CHIANTI is a collaborative project involving the
NRL (USA), the
Universities of Florence (Italy) and Cambridge (UK), and George Mason University
(USA). We are grateful for the open data policies of {\it Hinode} and the efforts 
of the instrument and software teams. Hinode is a Japanese mission developed and 
launched by ISAS/JAXA, with NAOJ as domestic partner and NASA and STFC (UK) as international partners. 
It is operated by these agencies in co-operation with ESA and NSC (Norway). 

\bibliographystyle{apj}
\bibliography{dem_bib}

\end{document}